%% file: EXO-14-009_temp.tex
\pdfoutput=1

\documentclass[11pt,twoside,a4paper,cmspaper,final,collab]{cms-tdr}
\def\svnVersion{342837}\def\cmsCernNoTag{CERN-PH-EP/2015-096}\def\cmsCernDate{\today}\def\cmsMessage{Published in the Journal of High Energy Physics as \href{http://dx.doi.org/10.1007/JHEP02(2016)145}{\doi{10.1007/JHEP02(2016)145.}}}

\begin{document}\cmsNoteHeader{EXO-14-009}

\hyphenation{had-ron-i-za-tion}
\hyphenation{cal-or-i-me-ter}
\hyphenation{de-vices}
\RCS$Revision: 318116 $
\RCS$HeadURL: svn+ssh://svn.cern.ch/reps/tdr2/papers/EXO-14-009/trunk/EXO-14-009.tex $
\RCS$Id: EXO-14-009.tex 318116 2016-01-18 13:10:10Z alverson $

\newcommand\Hbb{\ensuremath{\PH\to\bbbar}}
\newcommand\Hww{\ensuremath{\PH\to\PW\PW^*\to4{\PQq}}}
\newcommand\Zqq{\ensuremath{\cPZ\to\qqbar}}
\newcommand\Wqq{\ensuremath{\PW\to\cPaq \Pq'}}
\newcommand\Vqq{\ensuremath{\PW/\cPZ\to\cPaq \Pq'}}
\newcommand\HbbZqq{\ensuremath{\Hbb,\Zqq}}
\newcommand\HbbWqq{\ensuremath{\Hbb,\Wqq}}
\newcommand\HbbVqq{\ensuremath{\Hbb,\Vqq}}
\newcommand\HwwZqq{\ensuremath{\Hww,\Zqq}}
\newcommand\HwwWqq{\ensuremath{\Hww,\Wqq}}
\newcommand\HwwVqq{\ensuremath{\Hww,\Vqq}}
\newcommand{\ptk}{\ensuremath{p_{\mathrm{T},k}}\xspace}
\providecommand{\cPV}{\ensuremath{\mathrm{V}}\xspace}

\newcommand\HbbAll{\ensuremath{\mathrm{VH_{bb}}}}
\newcommand\HWWAll{\ensuremath{\mathrm{VH_{WW}}}}
\newcommand\HbbHP{\ensuremath{\mathrm{V^{HP} H_{bb}}}}
\newcommand\HbbLP{\ensuremath{\mathrm{V^{LP} H_{bb} }}}
\newcommand\HWWHP{\ensuremath{\mathrm{V^{HP} H_{WW}^{HP}}}}
\newcommand\HWWLPH{\ensuremath{\mathrm{V^{HP} H_{WW}^{LP}}}}
\newcommand\HWWLPV{\ensuremath{\mathrm{V^{LP} H_{WW}^{HP}}}}

\newcommand{\scalefactorHP}{ \ensuremath{0.86 \pm 0.07}\xspace}
\newcommand{\scalefactorLP}{ \ensuremath{1.39 \pm 0.75}\xspace}
\newcommand{\scalefactorHPu}{ \ensuremath{7.5\%}\xspace}
\newcommand{\scalefactorLPu}{ \ensuremath{54\%}\xspace}
\newcommand{\intlumi}{19.7\fbinv}

 \cmsNoteHeader{EXO-14-009}

\title{Search for a massive resonance decaying into
a Higgs boson and a W or Z boson in hadronic final states
in proton-proton collisions at $\sqrt{s}= 8$\TeV}

 \date{\today}

 \abstract{

  A search for a massive resonance decaying into a
  standard-model-like Higgs boson (H) and a W or Z boson is reported.
  The analysis is
  performed on a data sample
  corresponding to
  an integrated luminosity of 19.7\fbinv,  collected in
  proton-proton collisions at a centre-of-mass energy of 8\TeV with
  the CMS detector at the LHC.
  Signal events, in which the decay products of
  Higgs, W, or Z bosons at high Lorentz boost
   are contained within single
  reconstructed jets, are identified using jet substructure techniques,
  including the tagging of b hadrons.
  This is the first search for heavy resonances decaying into HW or HZ
  resulting in an all-jet final
  state, as well as the first application of jet substructure techniques
  to identify
  $\PH\to \PW\PW^*\to 4\PQq$ decays at high Lorentz boost.
  No significant signal is observed and
  limits are set at 95\% confidence level on the production cross sections of
  $\PWpr$ and $\cPZpr$ in a model with mass-degenerate charged and neutral spin-1
  resonances.
  Resonance masses are excluded
  for $\PWpr$ in the interval [1.0, 1.6]\TeV,
  for $\cPZpr$ in the intervals [1.0, 1.1] and [1.3, 1.5]\TeV,
  and for
  mass-degenerate $\PWpr$ and $\cPZpr$ in the interval [1.0, 1.7]\TeV.
}

\hypersetup{%
pdfauthor={CMS Collaboration},%
pdftitle={Search for a massive resonance decaying into a Higgs boson and a W or Z boson in hadronic final states in proton-proton collisions at sqrt(s) = 8 TeV},%
pdfsubject={CMS},%
pdfkeywords={CMS, physics, Higgs}}

\maketitle

 \section{Introduction}
\label{sec:introduction}

Several theories of physics beyond the standard model (SM) predict the
existence of vector resonances with
masses above 1\TeV that decay into
a W or Z vector boson (V)
 and a SM-like Higgs boson (H).
Here we present a search for
the production of such resonances
 in proton-proton (pp) collisions at a
centre-of-mass energy of $\sqrt{s}=8$\TeV.  The data sample,
corresponding to an integrated luminosity of 19.7\fbinv, was collected
with the CMS detector at the CERN LHC.

The composite Higgs~\cite{Composite1,Composite2, Composite0} and
little Higgs models~\cite{Han:2003wu, Schmaltz:2005ky, Perelstein:2005ka}
address the hierarchy problem and predict many new particles,
including additional gauge bosons, e.g. heavy spin-1 $\PWpr$ or $\cPZpr$ bosons ($\cPV'$).
These models can be generalized in the heavy vector triplet (HVT) framework~\cite{Pappadopulo:2014qza}.
Of particular interest for this search is the HVT scenario B model,
where the branching fractions $\mathcal{B}(\PWpr \to \PW\PH)$
and $\mathcal{B}(\cPZpr \to \Z\PH)$
dominate over the corresponding branching fractions to fermions,
and are comparable to  $\mathcal{B}(\PWpr \to \PW\Z)$ and
 $\mathcal{B}(\cPZpr\to\PW\PW)$.
In this scenario, experimental constraints from searches for boson decay channels are more stringent than those from fermion decay channels.
Several searches~\cite{Khachatryan:2014xja,
Aad:2014pha,Aad:2014xka,ATLASWWPAPER,Khachatryan:2014hpa}
for $\PWpr\to\PW\Z$
based upon the Extended Gauge Boson (EGB) reference
model~\cite{egm} have excluded resonance
masses below 1.7\TeV. Unlike the HVT scenario B model,
the EGB model has enhanced fermionic couplings
and the mass limit is not directly
comparable to this work. Model independent limits on
the cross section for the resonant
production $\ell\nu+\text{jets}$~\cite{EXO-13-009} can
be used to extract resonance mass
limits on the processes $\PWpr\to \PW\Z$
and $\cPZpr \to \PW\PW$ of
1.7\TeV and 1.1\TeV, respectively.
A search for $\cPZpr\to \Z\PH \to \PQq\PAQq\tau\tau$
was reported in Ref.~\cite{cms-HZ-tautaujet} and
interpreted in the context of HVT scenario model B; however,
no resonance mass limit could be set with the sensitivity achieved.
Finally, a recent search~\cite{Aad:2015yza} combining leptonic
decays of W and Z bosons, and two b-tagged jets forming a $\Hbb$ candidate
excluded HVT model A with coupling constant $g_V = 1$ for heavy vector
boson masses below $m_{\cPV'^0} < 1360\GeV$ and $m_{\cPV'^\pm} < 1470\GeV$.

The signal of interest is a narrow heavy vector resonance $\cPV'$ decaying into
VH, where the V decays to a pair of quarks and the H decays either to
a pair of b quarks, or to a pair of W bosons, which further decay into
quarks.
The H in the HVT framework does not have properties that are identical
to those of a SM Higgs boson. We make the assumption that the state
observed by the LHC Collaborations~\cite{higgsdiscoveryAtlas,Chatrchyan:2012ufa}
is the same as the one described by the HVT framework and
that, in accord with present
measurements~\cite{Khachatryan:2014kcaCMSHiggs1,Khachatryan:2014jbaCMSHiggs2},
its properties are similar to those of a SM Higgs boson.

In the decay of massive $\cPV'$ bosons
produced in the pp collisions at the LHC,
the momenta of the daughter V and H are large enough (${>}$200\GeV)
that their hadronic decay products
are reconstructed as single jets~\cite{Gouzevitch:2013qca}.
Because this results in a dijet topology,
traditional analysis techniques relying on resolved jets are
no longer applicable. The signal is characterized by a peak
in the dijet invariant mass ($m_\mathrm{jj}$) distribution
over a continuous background from mainly QCD multijet
events. The sensitivity to b-quark jets
from H decays is enhanced through
subjet or jet b tagging~\cite{BTV-13-001}.
Jets from $\Vqq$, $\Hbb$, and $\Hww$
decays are identified with jet
substructure techniques~\cite{topwtag_pas,JME-13-006}.

This is the first search for heavy resonances
decaying via VH into all-jet final states
and it incorporates the first application of jet substructure
techniques to identify $\Hww$ at a high Lorentz boost.

 \section{The CMS detector}
\label{sec:cms_detector}
The central feature of the CMS apparatus is a
 superconducting solenoid of 6\unit{m}
internal diameter, providing a magnetic
 field of 3.8\unit{T}. Within the field
volume are a silicon pixel and strip tracker, a lead
tungstate crystal electromagnetic calorimeter,
and a brass and scintillator hadron calorimeter,
 each composed of a barrel and two endcap sections.
Muons are measured in gas-ionization detectors embedded
 in the steel flux-return yoke outside the solenoid.
Extensive forward calorimetry complements the coverage
provided by the barrel and endcap detectors.
A more detailed description of the CMS detector,
together with a definition of the coordinate system used and
the relevant kinematic variables,
can be found in Ref.~\cite{Chatrchyan:2008zzk}.

 \section{Signal model and simulation}
\label{sec:signalModel}

In the HVT framework, the production cross sections of
$\PWpr$ and $\cPZpr$ bosons and their decay branching fractions depend on
three parameters in addition to the resonance masses:  the
strength of couplings to quarks ($c_\PQq$), to the H ($c_\PH$),  and on
their self-coupling ($g_{\cPV}$).
In the HVT model B,
where $g_{\cPV}=3$ and $c_{\PQq}=-c_{\PH}=1$,
 $\PWpr$ and $\cPZpr$ preferentially couple to bosons (W/Z/H),
giving rise to diboson final states.
This feature reproduces the properties of the $\PWpr$ and $\cPZpr$ bosons
 predicted by the minimal composite Higgs model.
In this case, the production cross sections for \cPZpr,
${\PWpr}^{-}$, and ${\PWpr}^{+}$ are respectively 165,
87, and 248\unit{fb}
for a signal of resonance mass $m_{\cPV'} =1\TeV$.
Their branching
fractions to VH and decay width are respectively
51.7\%, 50.8\%, 50.8\% and 35.0, 34.9, 34.9\GeV.
The resonances are assumed to be narrow, \ie, with natural
widths smaller than the experimental resolution in $m_\mathrm{jj}$ for
masses considered in this analysis.

We consider the $\PWpr$ and $\cPZpr$
resonances separately, and report limits for
each candidate individually to permit
the reinterpretation of our results in
different scenarios with different numbers of spin-1 resonances.

Signal events are simulated using
the \MADGRAPH~5.1.5.11~\cite{madgraph} Monte Carlo event
generator to generate partons that are
then showered with \PYTHIA~6.426~\cite{Sjostrand:2006za} to
produce final state particles. These events are then
processed through a \GEANTfour~\cite{refGEANT} based simulation of
the CMS detector. The \MADGRAPH input parameters are
provided in Ref.~\cite{Pappadopulo:2014qza1} and the H mass
is assumed to be 125\GeV. Samples showered
with \HERWIG{++}~2.5.0~\cite{herwig} are
used to evaluate the systematic uncertainty
associated with the hadronization. Tune Z2*~\cite{bib_tunez1} is
used in \PYTHIA, while the version 23 tune~\cite{herwig} is
used in \HERWIG{++}. The CTEQ6L1~\cite{cteq} parton distribution
functions (PDF) are used for \MADGRAPH, \PYTHIA and \HERWIG{++}.
Signal events are generated from resonance mass 1.0 to 2.6\TeV in
steps of 0.1\TeV. Signals with resonance masses
between the generated values are interpolated.

The distribution of the background is modelled from the data.
However, simulated samples of multijet and $\ttbar$ events, generated
using {\MADGRAPH 5v1.3.30}~\cite{madgraph} and {\POWHEG
1.0}~\cite{Nason:2004,Frixione:2007,Alioli:2010xd}, respectively, and
interfaced to \PYTHIA for parton showering and hadronization, serve to
provide guidance and cross-checks.

\section{Event reconstruction and selection}\label{sec:analysis}

The event selection,
in the online trigger as well as offline,
utilizes a global event description
by combining information
from the individual subdetectors.
Online, events are
selected by at least one of two specific triggers:
 one based on the scalar sum of the transverse
momenta \pt of the jets ($\HT$),
which requires $\HT > 650\GeV$; the
other on the invariant mass
 of the two jets with highest \pt ,
which requires $m_\mathrm{jj} > 750\GeV$.

The offline reconstruction is described below.

Events must have at least
one primary vertex reconstructed
with $\abs{z} <
24\unit{cm}$. The primary vertex used in the event reconstruction is the one with the
largest summed $\pt^2$ of associated tracks. Individual particle
candidates
are reconstructed and identified using the particle-flow
algorithm~\cite{particleflow,particleflow2}, and divided into five
categories: muons, electrons, photons (including those that convert
into $\Pep\Pem$ pairs), charged hadrons, and neutral hadrons. Charged
particle candidates associated with a primary vertex different from
the one considered for the event reconstruction are discarded,
 which reduces contamination from additional pp interactions in the
same bunch crossings (pileup).

Jets are clustered from the remaining particle flow candidates, except those identified as isolated muons, using the
Cambridge--Aachen (CA)~\cite{CAaachen,CAcambridge} jet clustering
algorithm as implemented in
\FASTJET~\cite{fastjet1,fastjet}.
This algorithm starts from a set of particles as  ``protojets''.
It combines them iteratively with each other into new protojets
until the distance of the resulting protojet to the closest remaining
protojet is larger than the distance parameter of the CA algorithm.  A distance parameter of 0.8
is used (CA8 jets). An event-by-event correction based on
the jet area
method~\cite{jetarea_fastjet,jetarea_fastjet_pu,JME-JINST} is applied
to remove the remaining energy deposited by neutral particles
originating from pileup. The pileup-subtracted jet
four-momenta are then corrected to account for the difference between
the measured and true energies of hadrons~\cite{JME-JINST}.
Jet identification criteria~\cite{CMS-PAS-JME-10-003} are
applied to the two highest \pt jets
in order to remove spurious events associated with calorimeter noise.

The jet reconstruction
efficiencies (estimated from simulation) are larger than 99.9\%, and
contribute negligibly to the systematic uncertainties for signal
events.

Events are selected by requiring at least two jets each
with $\pt > 30\GeVc$ and
pseudorapidity $\abs{\eta} < 2.5$. The two highest $\pt$ jets are
required to have a pseudorapidity separation $\abs{\Delta\eta}<1.3$ to
reduce background from multijet events~\cite{cmsdijet}. The
invariant mass of these two jets is required to satisfy
$m_\mathrm{jj}> 890\GeVcc$.
The trigger efficiency for the events passing the preselection
requirements exceeds 99\%.

\begin{figure}[th!b]
\begin{center}
 \includegraphics[width=0.69\textwidth]{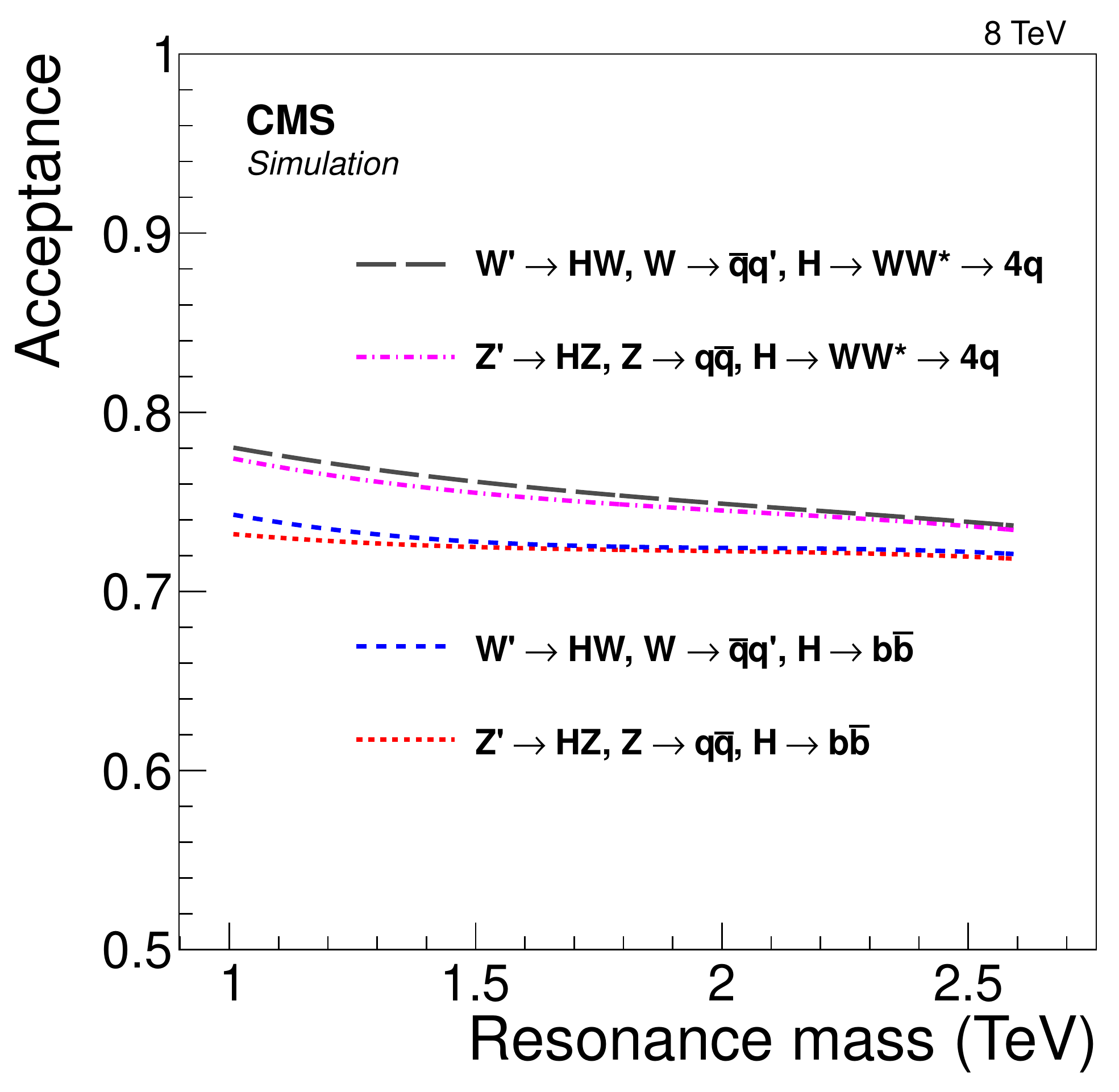}
 \end{center}
\caption{The fraction of simulated signal events for hadronically
  decaying W/Z and H
  bosons, reconstructed as two jets, that
  pass the geometrical acceptance criteria ($\abs{\eta} < 2.5$,
  $\abs{\Delta\eta}<1.3$), shown as a function of the resonance
  mass.
 }
\label{fig:acceptance}
\end{figure}

\begin{figure}[ht!b]
\begin{center}
\includegraphics[width=0.49\textwidth]{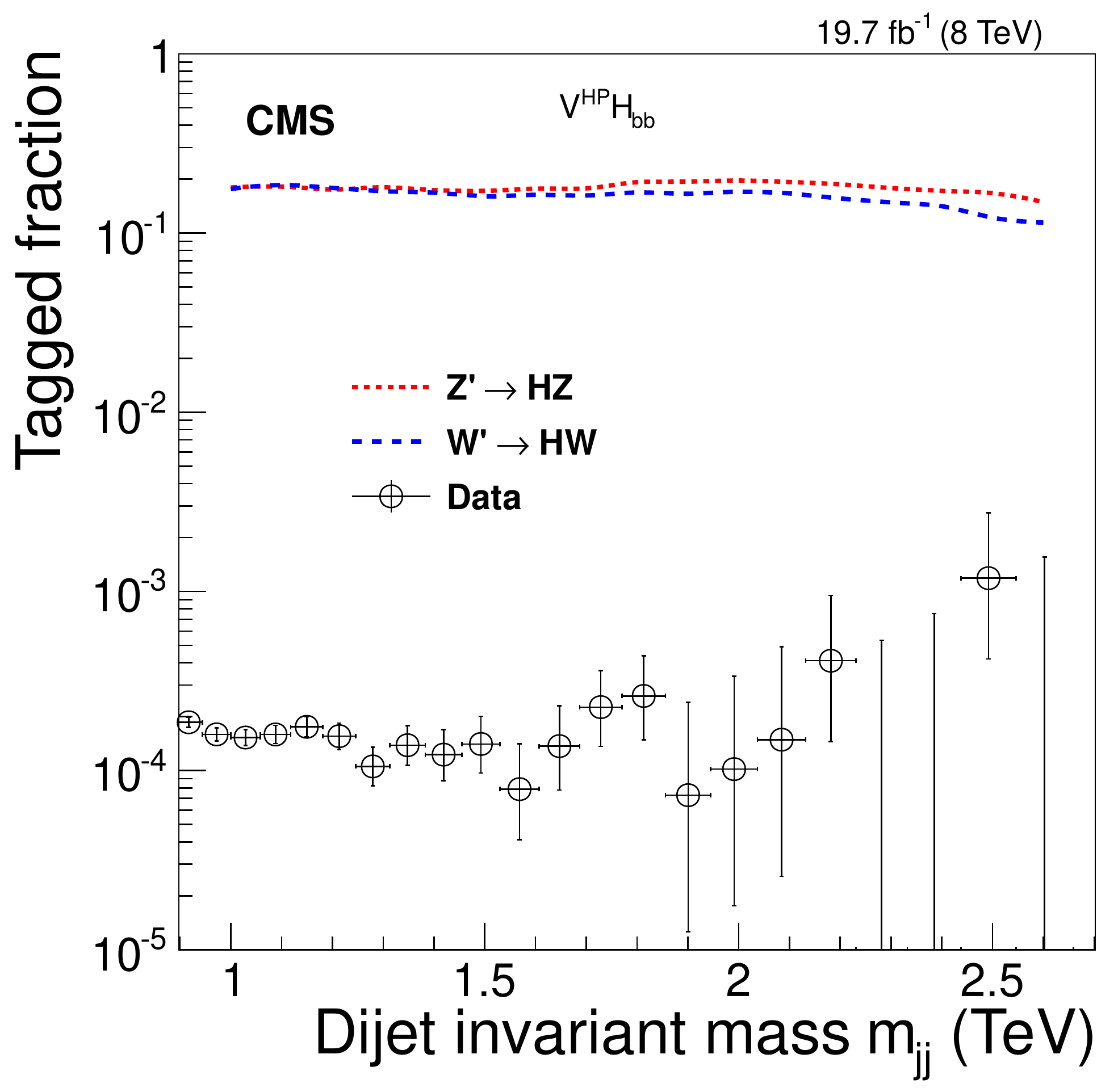}
\includegraphics[width=0.49\textwidth]{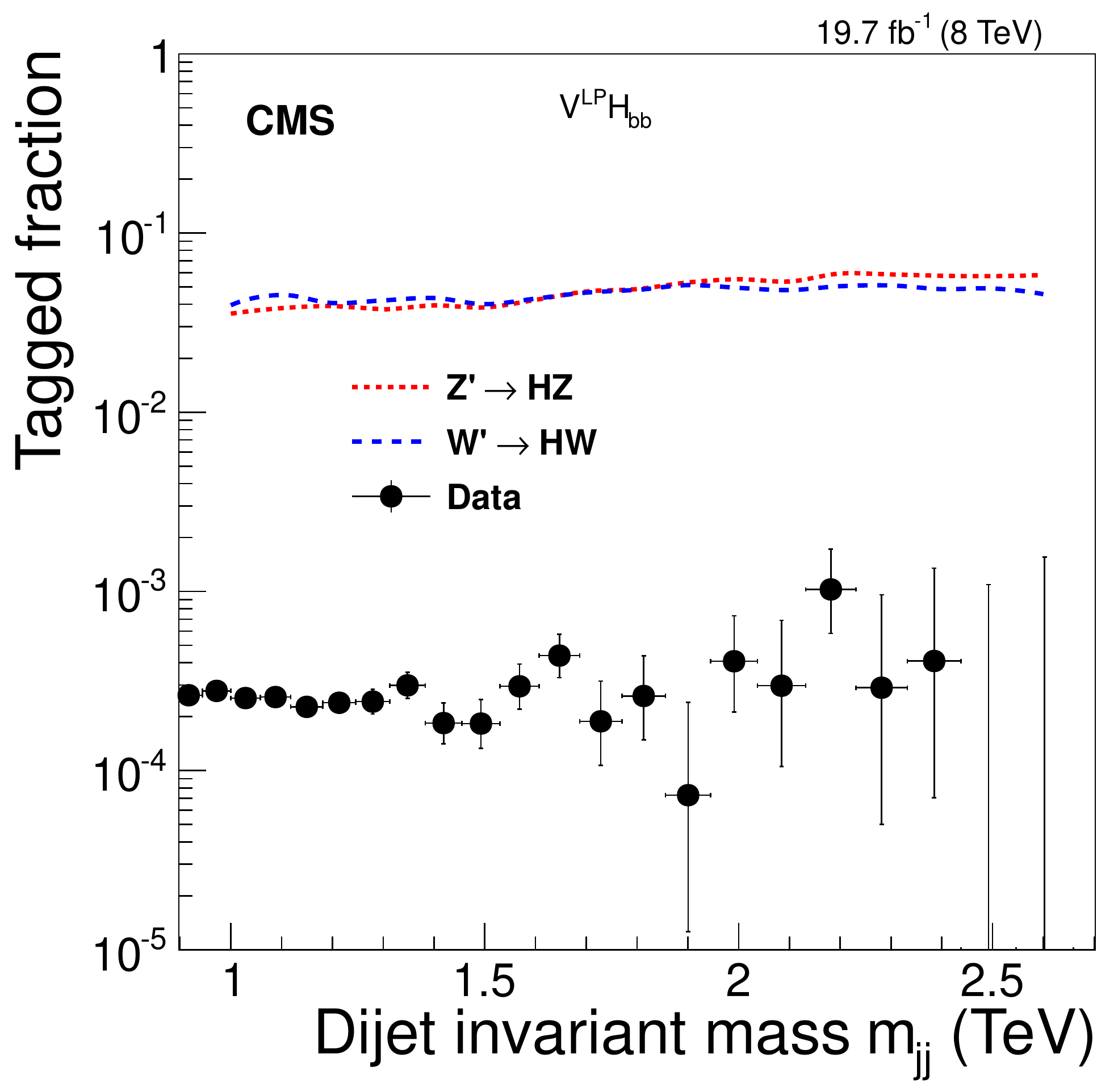}
\end{center}
\caption{
Tagged fractions in \HbbVqq\ signal channels and data
as a function of dijet invariant mass, for categories of
\HbbHP\ (left) and \HbbLP\ (right). Horizontal bars
through the data points indicate the bin width.
}
\label{fig:HbbZqqOverallEff}
\end{figure}

\begin{figure}[ht!b]
\begin{center}
\includegraphics[width=0.51\textwidth]{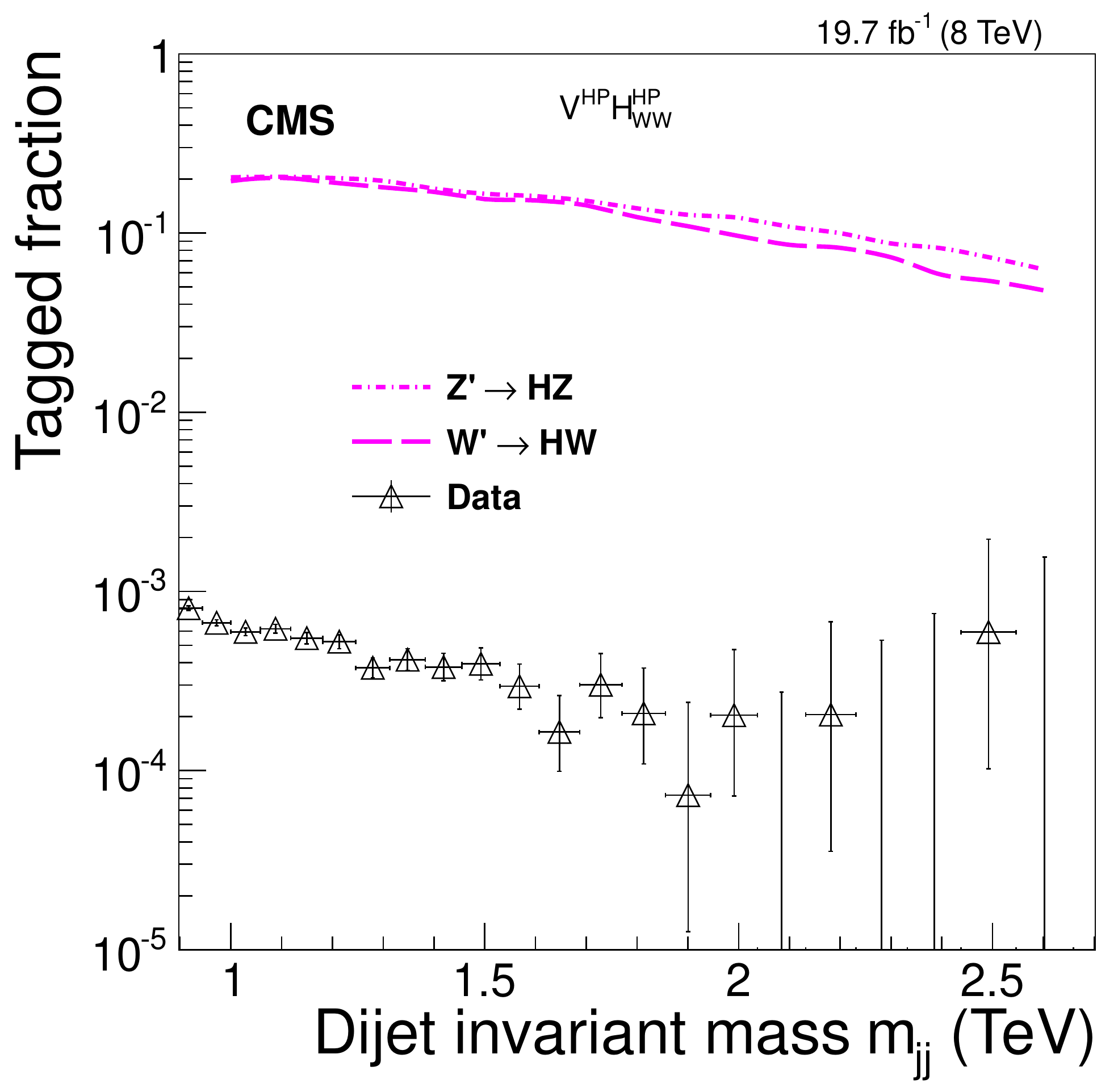}
\includegraphics[width=0.49\textwidth]{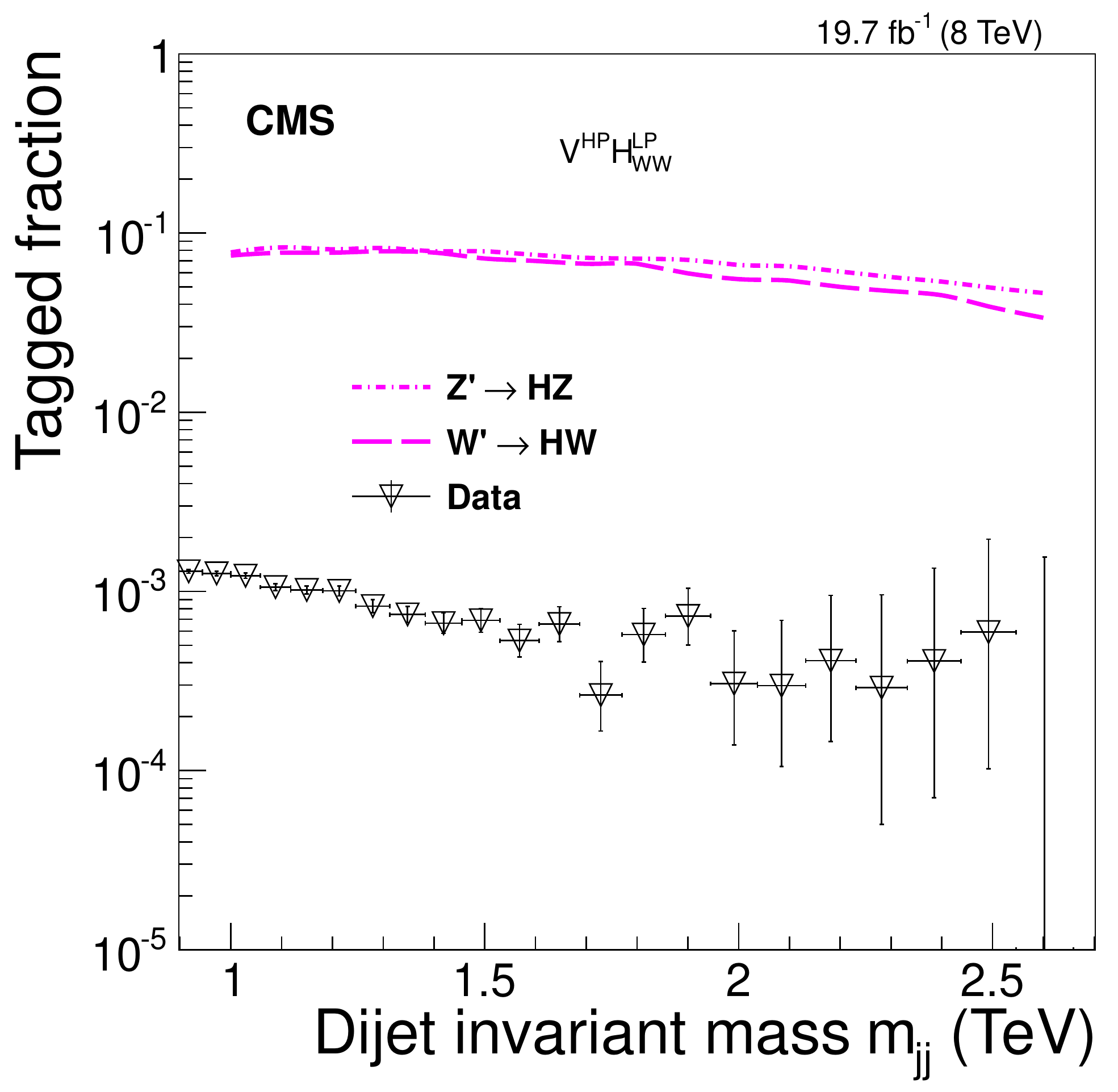}
\includegraphics[width=0.49\textwidth]{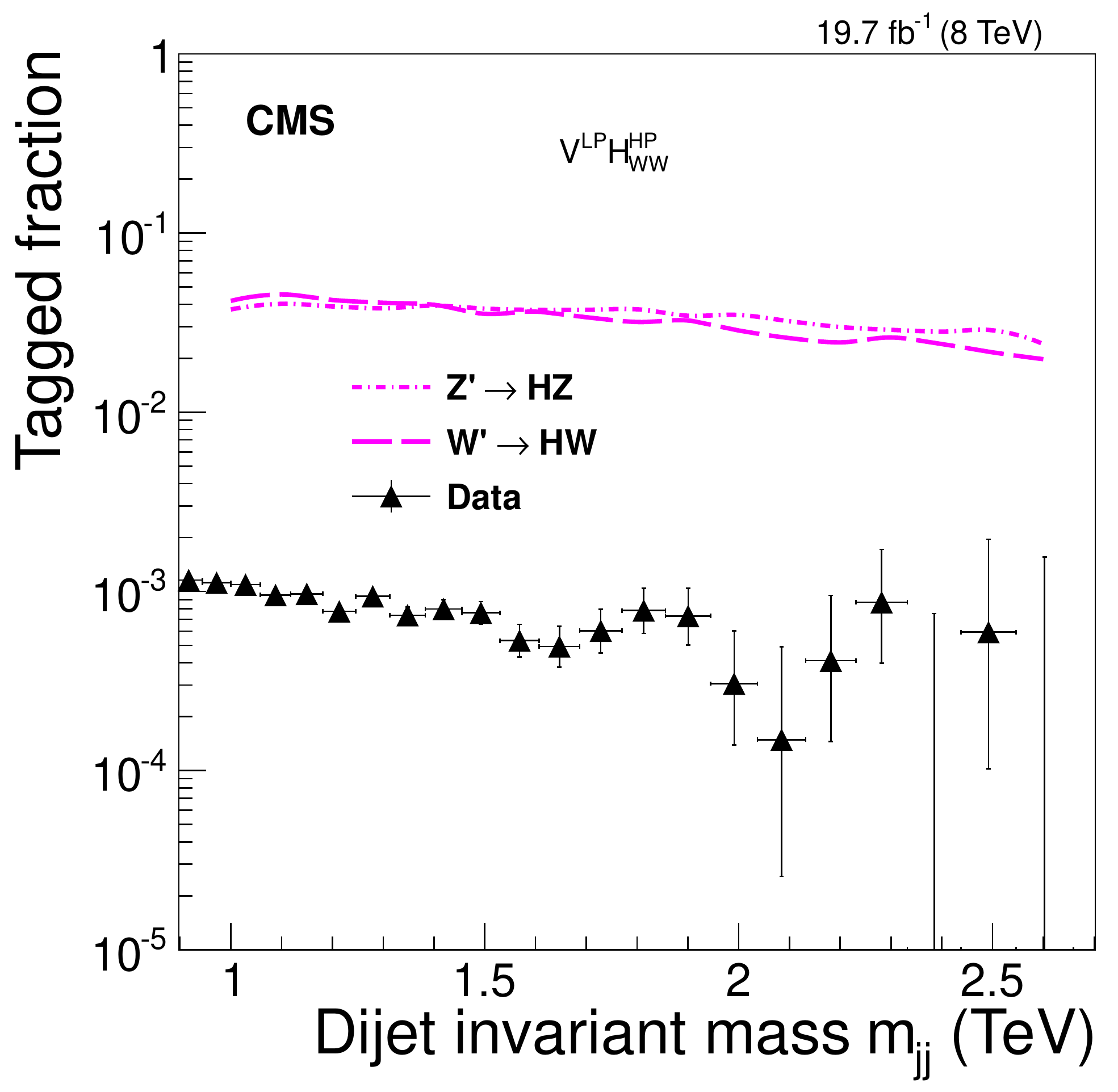}
\end{center}
\caption{
Tagged fractions in \HwwVqq\ signal channels and data as
a function of dijet invariant mass, for categories of
\HWWHP\ (top), \HWWLPH\ (bottom left) and \HWWLPV\ (bottom right).
Horizontal bars
through the data points indicate the bin width.
}
\label{fig:HwwEffAll}
\end{figure}

 To enable the results
to be applied to other models of similar final states,
we utilize simulations to derive the geometrical acceptances
and the W/Z and H selection efficiencies. These are presented separately in
Figs.~\ref{fig:acceptance},~\ref{fig:HbbZqqOverallEff}, and~\ref{fig:HwwEffAll}, respectively.

For the purpose of reinterpreting the result, the global efficiency is presented
approximated by the product of acceptances and
the W/Z and H selection efficiency, restricted
to final states where the W/Z and H bosons decay
hadronically.
The products of acceptances and the W/Z and H tagging efficiency,
ignoring the correlations between
detector acceptance and W/Z or H tagging,
agree to better than 10\% with the full event simulation.
In the interpretations reported in this paper,
the global efficiency is estimated from the full simulation
of signal events, such that the correlations
between the acceptance and W/Z and H selection efficiency
are properly taken into account.
However, when re-interpreting this search in terms of an arbitrary
model, an additional uncertainty of 10\% should be folded in, to allow
for the possible effect of correlations.

The acceptance, shown in Fig.~\ref{fig:acceptance}
as a function of the dijet resonance mass for several
signals, takes into account the angular
acceptance ($\abs{\eta} < 2.5$, $\abs{\Delta\eta}<1.3$).

The two highest $\pt$ jets are chosen as candidates for the
hadronically decaying W/Z and H
bosons, and W/Z and H tagging algorithms
 based on jet substructure are applied.

Information characterizing jet substructure is derived using three
separate algorithms, producing the
variables \textit{pruned jet mass}, \textit{subjet b tagging}, and \textit{N-subjettiness}.
The combined use of these variables
in event selection strongly suppresses the background from QCD dijet
production.  All three characterizations of jet substructure are defined
and discussed in detail in the following paragraphs.

\begin{figure}[ht!b]
\begin{center}
\includegraphics[width=0.69\textwidth]{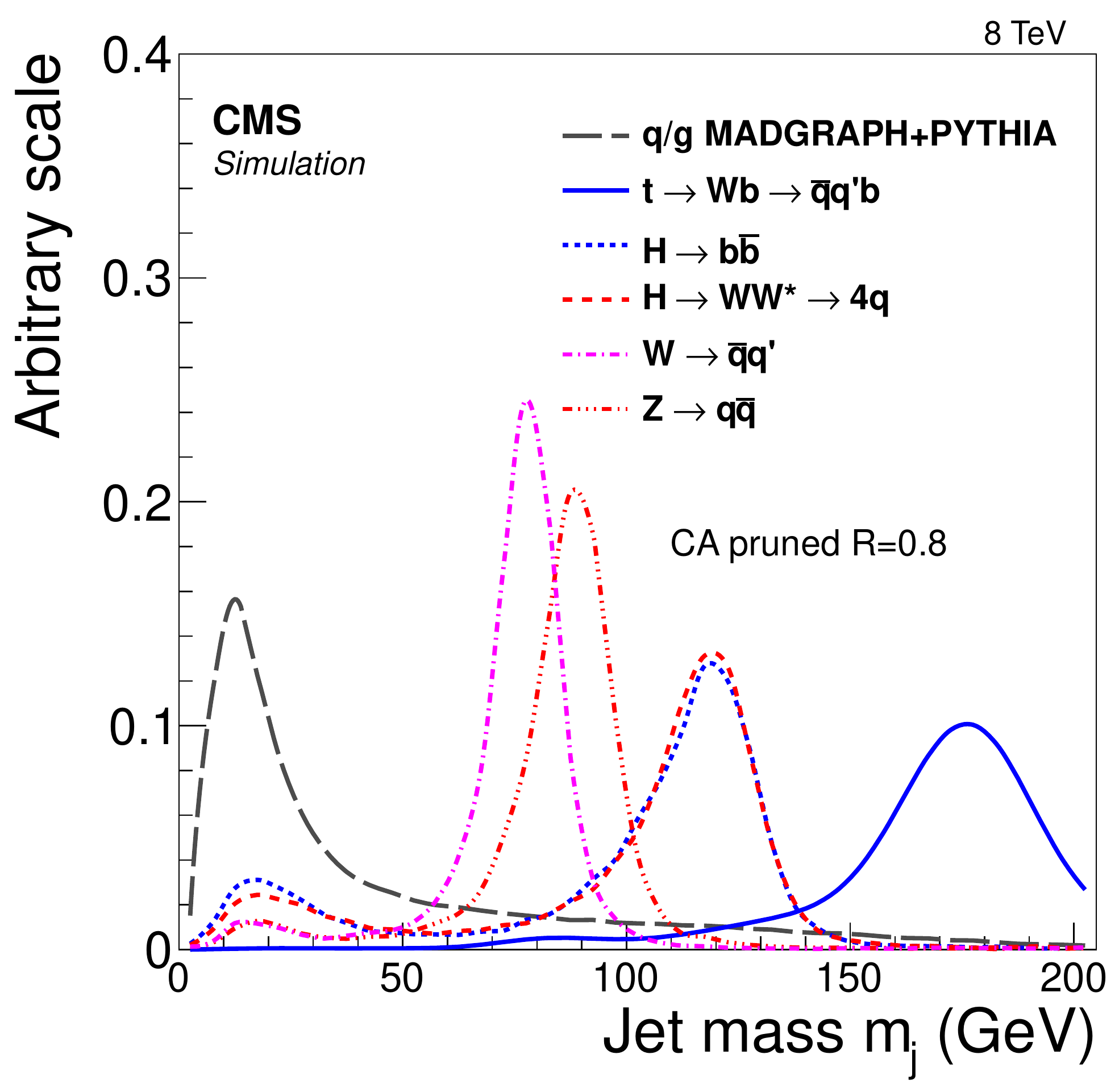}
\end{center}
\caption{Distribution of pruned jet mass in
  simulation of signal and background processes.
  All simulated distributions are normalized to 1.
  The W/Z, H, and top-quark jets are required to
  match respective generator level particles in the event.
  The W/Z and H jets are from 1.5\TeV $\PWpr\to \PW\PH$ and
  $\cPZpr \to \Z\PH$ signal samples.
}
\label{fig:JetMassTagging}
\end{figure}

As the mass of the V or H boson
is larger than the mass of a typical QCD jet, the jet mass is the
primary observable that distinguishes such a jet from a QCD jet.  The
bulk of the V or H jet mass arises from the kinematics of the two or
more jet cores that correspond to the decay quarks.  In contrast, the
QCD jet mass arises mostly from soft gluon radiation.  For this
reason, the use of jet pruning~\cite{jetpruning1,jetpruning2} improves
discrimination by removing the softer radiation, as this shifts the
jet mass of QCD jets to smaller values, while maintaining the jet mass
for V and H jets close to the masses of W, Z or H bosons.
Jet pruning is implemented by applying additional cuts in the process
of CA jet clustering.  These cuts remove
protojets that would have a large angle and low $\pt$
with respect to the combination with another protojet.
The details of this procedure are given in Ref.~\cite{JME-13-006}.
 The distributions of the pruned jet mass ($m_\mathrm{j}$) for
simulated signal and background samples, are shown in
Fig.~\ref{fig:JetMassTagging}.  Jets from boosted \PW\ and
\cPZ\ decays are expected to generate peaks at $m_\mathrm{j} \approx
80$ and $m_\mathrm{j} \approx 90\GeV$, respectively. Jets
from boosted H decays are expected to peak at
$m_\mathrm{j} \approx 120\GeV$.
Hadronic top-quark jets, where the b quark and the
two different light quarks from the $\PQt \to \PW\PQb \to \PQq\PAQq'\PQb$ decay are required
to be within a reconstructed CA8 jet,
peak at $m_\mathrm{j} \approx 175\GeV$.
The peak around 20\GeV arises from unmerged light jets,
mostly associated with quark- and gluon-induced jets from multijet events,
but also from quark jets from W, \cPZ, and H bosons in the cases where the
decay products
do not end up in a single jet.  The contribution from bosons
depends on their spin and polarization.
All peaks are slightly shifted to lower masses because of the removal of soft
radiation in jet pruning.
If the pruned jet has a mass ($m_\mathrm{j}$)
within $70 <m_\mathrm{j}<100\GeVcc$ ( $110 <m_\mathrm{j}<135\GeVcc$ ), it is tagged as a $\PW/\cPZ$ ( H )
candidate.

Jet pruning can also provide a good delineation of subjets within the CA8 jet.

To tag jets from $\Hbb$ decays, denoted as $\PH_{\PQb\PQb}$ jets, the pruned subjets,
given by reversing the last step of the CA8 pruning recombination algorithm,
are used as the basis for b tagging.
Jets arising from the hadronization of b quarks (b jets) are identified using the ``Combined
Secondary Vertex'' b-tagging algorithm~\cite{CSVBtagging}, which uses information from tracks and
secondary vertices associated with jets to build a likelihood-based discriminator to distinguish
between jets from b quarks and those from charm or light quarks and gluons. The b-tagging
discriminator can take values between 0 and 1 with higher values indicating higher probability
for the jet to originate from a b quark.
The ``loose'' working point of the b-tagging algorithm~\cite{CSVBtagging}
is chosen and is found to be optimal
for both subjet and jet b tagging.
It has a b-tagging efficiency of ${\approx}85\%$,
with mistagging probabilities of ${\approx}40\%$
for c-quark jets and ${\approx}10\%$ for
light-quark and gluon jets at jet \pt near 80\GeV.
The ratio of b-tagging efficiencies
for data and simulation is applied as
a scale factor~\cite{BTV-13-001} to the simulated signal events.
To identify CA8 jets originating from $\Hbb$ decays,
we apply b tagging either to the two subjets or to the CA8 jet,
based on the angular separation of the two
subjets ($\Delta R$)~\cite{BTV-13-001}.
If $\Delta R$ is larger (smaller) than 0.3,
the b-tagging algorithm is applied
to both of the subjets (the single CA8 jet).

While the pruned jet mass is
a powerful discriminant against QCD multijet backgrounds,
the substructure of jets arising from V and H decays provides
additional discrimination. In $\Hww$ decays, the boosted H
decays into a final state of four quarks merged together,
denoted as an $\PH_{\PW\PW}$ jet,
and has a different substructure
than jets from $\cPV/\PH\to\PAQq\PQq'$ decays.
We quantify how well the constituents of
a given jet can be arranged into $N$ subjets
by reconstructing the full set of jet
constituents (before pruning) with
the \kt algorithm~\cite{ktalg} and halting the reclustering
when $N$ distinguishable protojets are formed.
The directions of the
$N$ jets are used as the reference axes to compute the
$N$-subjettiness~\cite{Thaler:2010tr,Thaler:2011gf,Stewart:2010tn}
$\tau_{N}$ of the original jet, defined as
\begin{equation}
\label{eq:subjettiness}
\tau_N = \frac{1}{d_{0}} \sum_{k} \ptk\,\min( \Delta R_{1,k}, \Delta R_{2,k},\ldots,\Delta R_{N,k}),
\end{equation}
where $\ptk$ is the $\pt$ of the $k^{th}$ constituent
of the original jet and $\Delta R_{n,k}$ is its angular distance from
the axis of the $n$th subjet (with $n=1, 2,\ldots,N$). The
normalization factor $d_{0}$ for $\tau_N$ is
$d_{0}= \sum_{k}\ptk R_{0}$, with $R_{0}$ set to $0.8$, the distance
parameter of the CA algorithm.
To improve the discriminating power, we perform a
one-pass optimization of the directions of the subjets' axes by
minimizing $\tau_{N}$~\cite{Thaler:2011gf,JME-13-006}.
By using the smallest $\Delta R_{n,k}$ to weight the value of $\ptk$ in
Eq.~(\ref{eq:subjettiness}), $\tau_N$ yields small values
when the jet originates from the hadronization of $N$ or fewer quarks.
 The $\tau_{ij} = \tau_{i} / \tau_{j}$
ratios $\tau_{21}$, $\tau_{31}$,
$\tau_{32}$, $\tau_{41}$, $\tau_{42}$, and $\tau_{43}$
have been studied to identify
the best discriminators for jets from $\Vqq$ and $\Hww$ decays.

\begin{figure}[ht!b]
\centering
\includegraphics[width=0.69\textwidth]{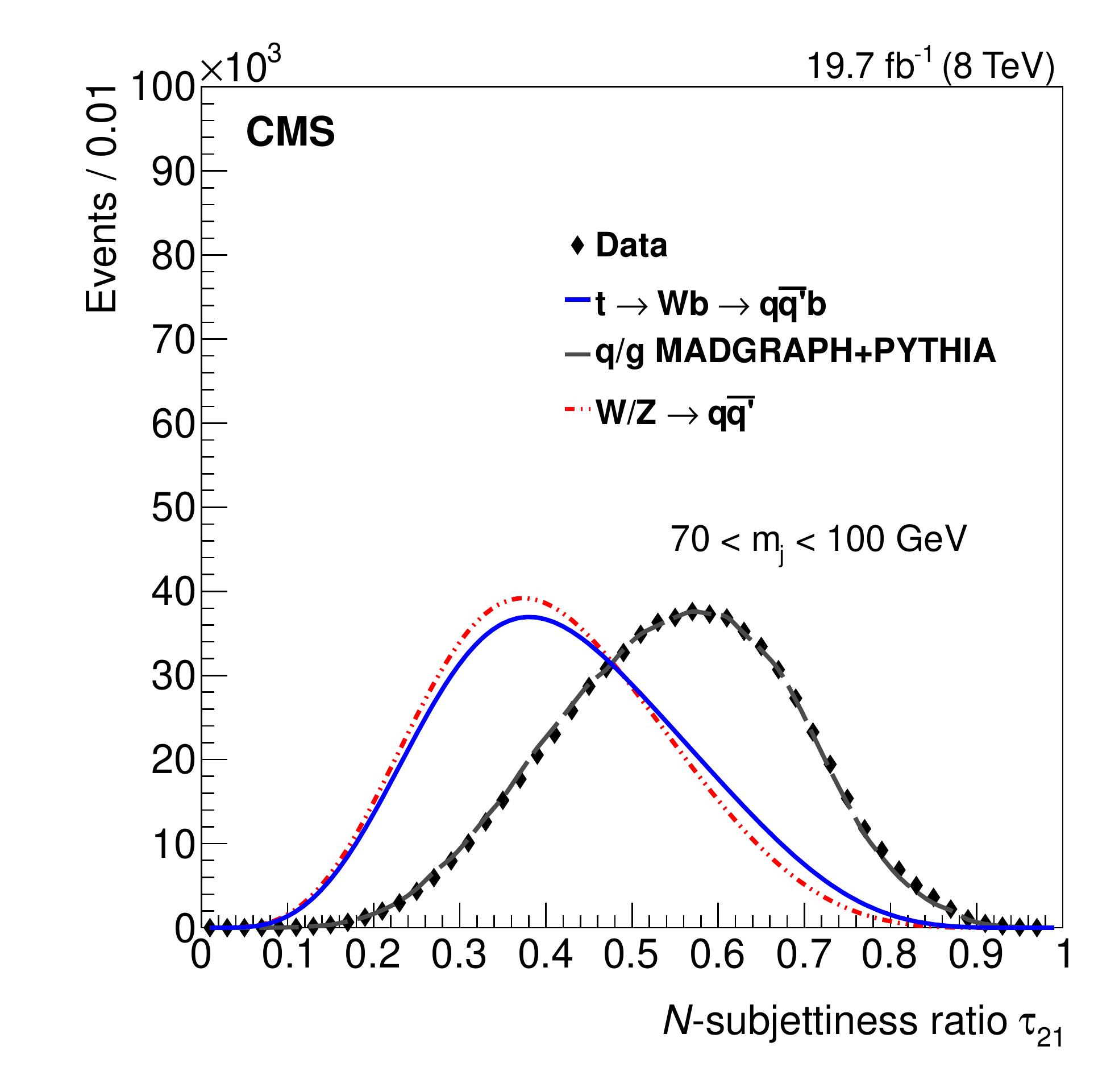}
\caption{Distribution of the $N$-subjettiness ratio
  $\tau_{21} = \tau_2/\tau_1$, where $\tau_N$ is
  given in Eq.~(\ref{eq:subjettiness}), for simulated
  signal and background processes, and for data.  The jets for
  which $\tau_{21}$ is calculated are required to satisfy the
  W/Z pruned jet mass requirement.
  The W/Z and top-quark jets are required to
  match respective generator level particles in the event.
  All simulated distributions are scaled to the number
  of events in data.
}
\label{fig:tau21}
\end{figure}

We find that $\tau_{21}$ is the most suitable variable
for identifying $\Vqq$
jets~\cite{Khachatryan:2014hpa}.  The distribution of $\tau_{21}$ for
the $\Vqq$ signal, shown in
Fig.~\ref{fig:tau21},
peaks below $0.4$ and is almost fully contained within
$\tau_{21} <0.75$, where we place our cut. In contrast, the QCD background peaks
around $0.6$.  The figure shows only W/Z candidate jets with the pruned jet
mass in the W/Z boson mass window.  For this reason, the jets matched to the top quark
are mostly true W bosons, and appear signal-like.
However, they represent only a small fraction of the top quarks from
$\ttbar$ events (cf.~Fig.~\ref{fig:JetMassTagging}), since in the kinematic regime considered in this search,
the top quarks are highly boosted and the b jet rarely fails to merge
with the W jet.  The overall contribution from $\ttbar$, after the
full selection, is 1--3\%.

\begin{figure}[ht!b]
\centering
\begin{tabular}{cc}
     \resizebox{0.5\linewidth}{!}{\includegraphics{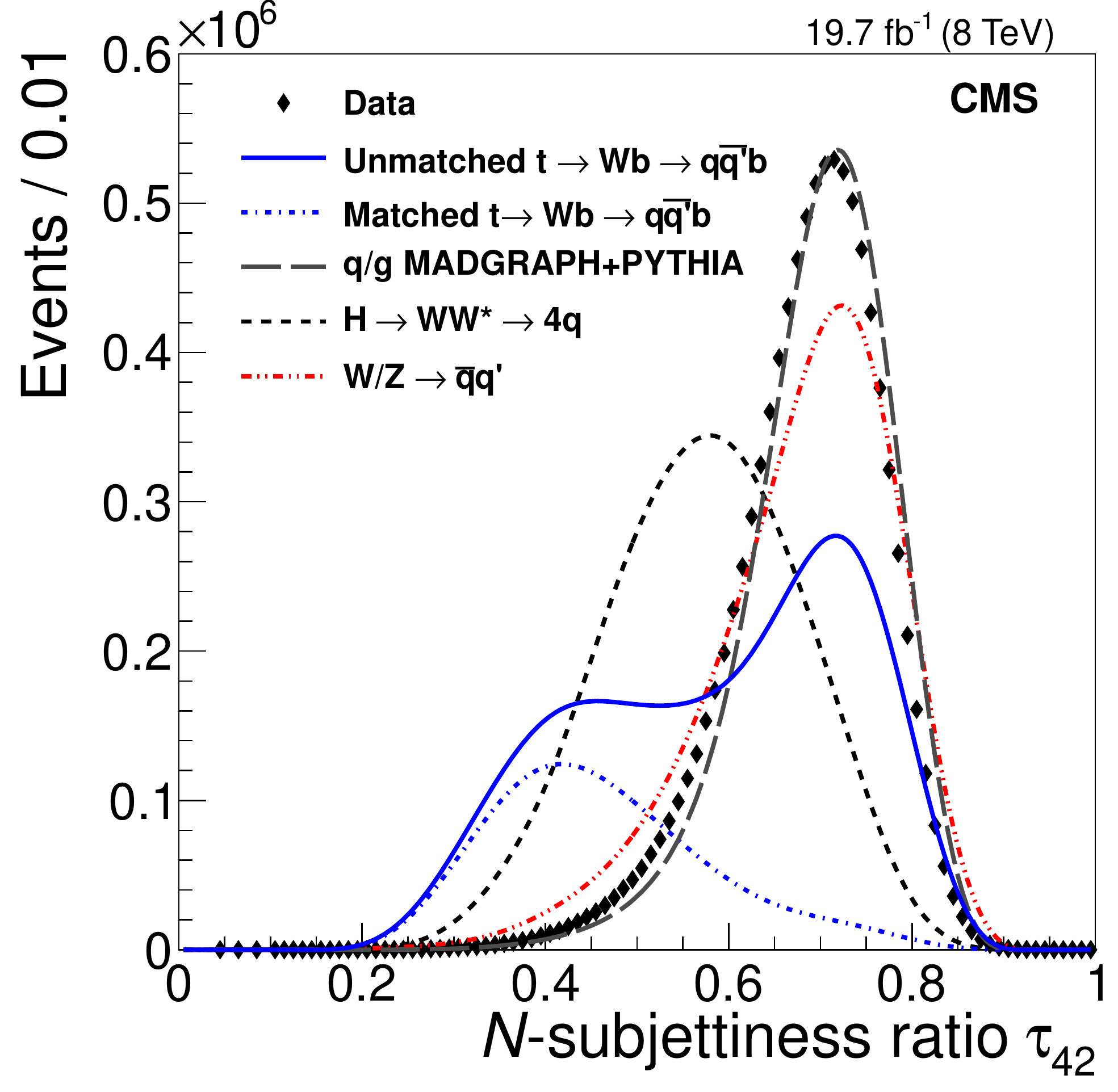}} &
     \resizebox{0.5\linewidth}{!}{\includegraphics{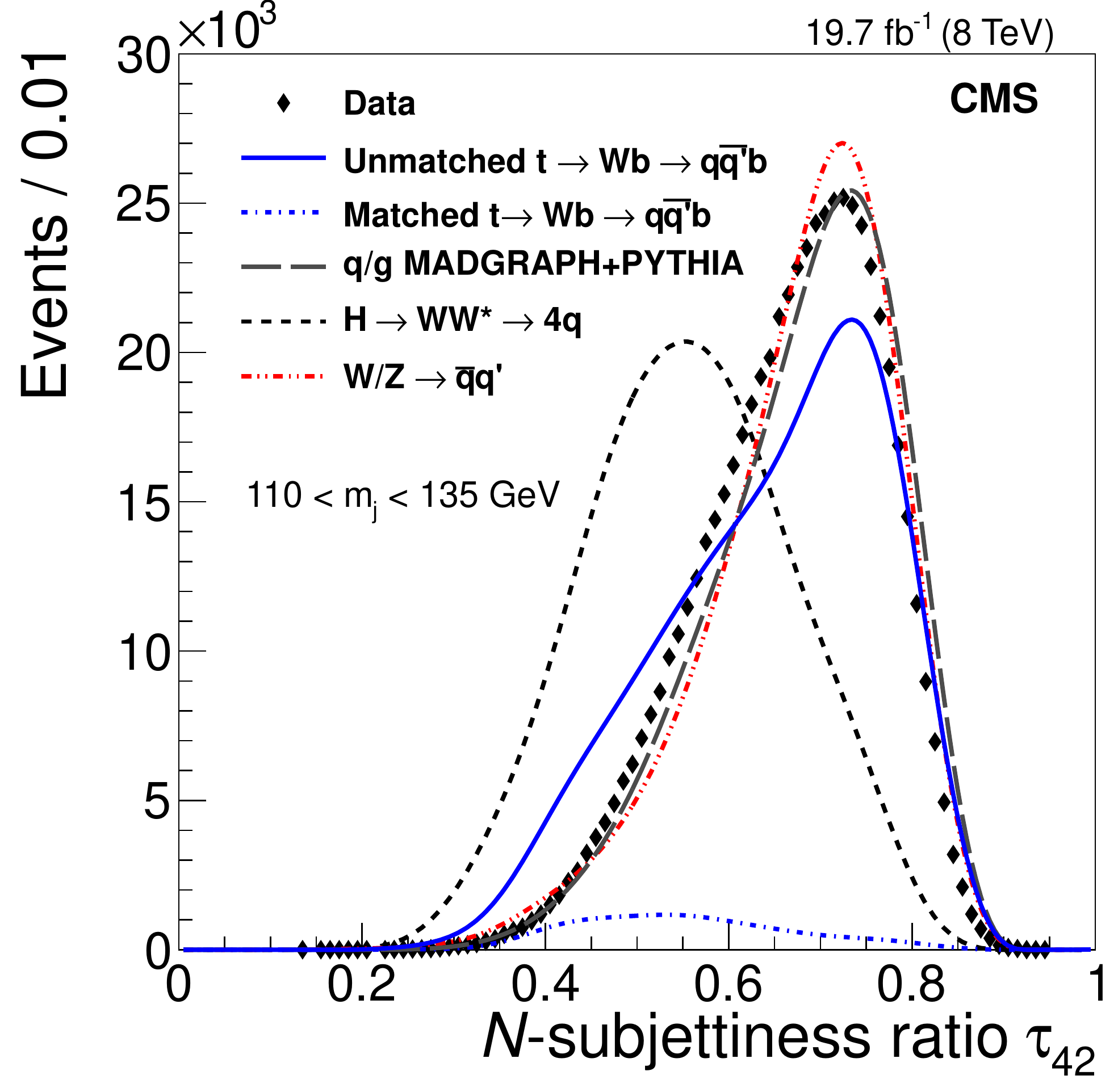}} \\
\end{tabular}
\caption
{
  Distributions of $\tau_{42}$ in
  data and in simulations of signal (2\TeV) and background events, without applying
  the pruned jet mass requirement (left)
  and with the pruned jet mass requirement applied (right).
  Matched top-quark, W/Z, and $\PH_{\PW\PW}$ jets are required to
  be consistent with their generator level particles, respectively.
  All simulated distributions are scaled to the number
  of events in data, except that matched top-quark background is scaled to
  the fraction of unmatched $\ttbar$ events times
  the number of data events.
  }
\label{fig:tau422TeV}
\end{figure}

For $\Hww$ events, we find that the ratio $\tau_{42}$
works best to
discriminate between four-pronged $\Hww$ and QCD jets.
The discriminating power of $\tau_{42}$ can
be seen in Fig.~\ref{fig:tau422TeV}.
The $\tau_{42}$ distribution of $\PH_{\PW\PW}$ jets tends to peak around 0.55.
By contrast, $\tau_{42}$ distributions of multijet background and W/Z jets
have a larger fraction of events at
large values of $\tau_{42}$, especially after requiring
a pruned jet mass in
the range [110, 135]\GeV.
Jets from unmatched $\ttbar$ events
peak together with QCD jets,
since they contain a mixture of b-quark jets and W-jets,
but relatively few fully merged top-quark jets.
However, the $\tau_{42}$ distribution for matched top-quark jets tends
to peak at smaller values, since for the same jet
$\tau_{42}$ is nearly always less than $ \tau_{32}$, which is
small for hadronic top-quark jets.

 In Fig.~\ref{fig:tau422TeV},
the comparison between dijet data and the QCD multijet simulation
shows that the simulated distribution
is well reproduced, though shifted towards
 higher values of $\tau_{42}$
as compared with the data.
A similar level of disagreement is
known for the modelling of $\tau_{21}$
in QCD simulation in Ref.~\cite{Khachatryan:2014hpa}.
The disagreement
does not affect this analysis
since the background is estimated from data.
 For the signal scale factor, the uncertainties from the
 modelling of $\tau_{42}$ are taken into account.

We select ``high (low)-purity'' $\PW/\cPZ$ jets by requiring $\tau_{21}
\leq 0.5$ ($ 0.5 < \tau_{21} < 0.75$), denoted
as the HP (LP) V tag.  Given the shape of $\tau_{21}$ distribution
for the  $\PW/\cPZ$ signal, the HP V tag category has a higher efficiency than
the LP V tag category.
We select HP (LP) $\PH_{\PW\PW}$ jets by requiring
$\tau_{42} \leq 0.55$ ($ 0.55 < \tau_{42} < 0.65$),
denoted as the HP (LP) H tag.  Here also the HP category has a higher
efficiency than the LP category.

\begin{figure}[ht!b]
\centering
\begin{tabular}{cc}
     \resizebox{0.49\linewidth}{!}{\includegraphics{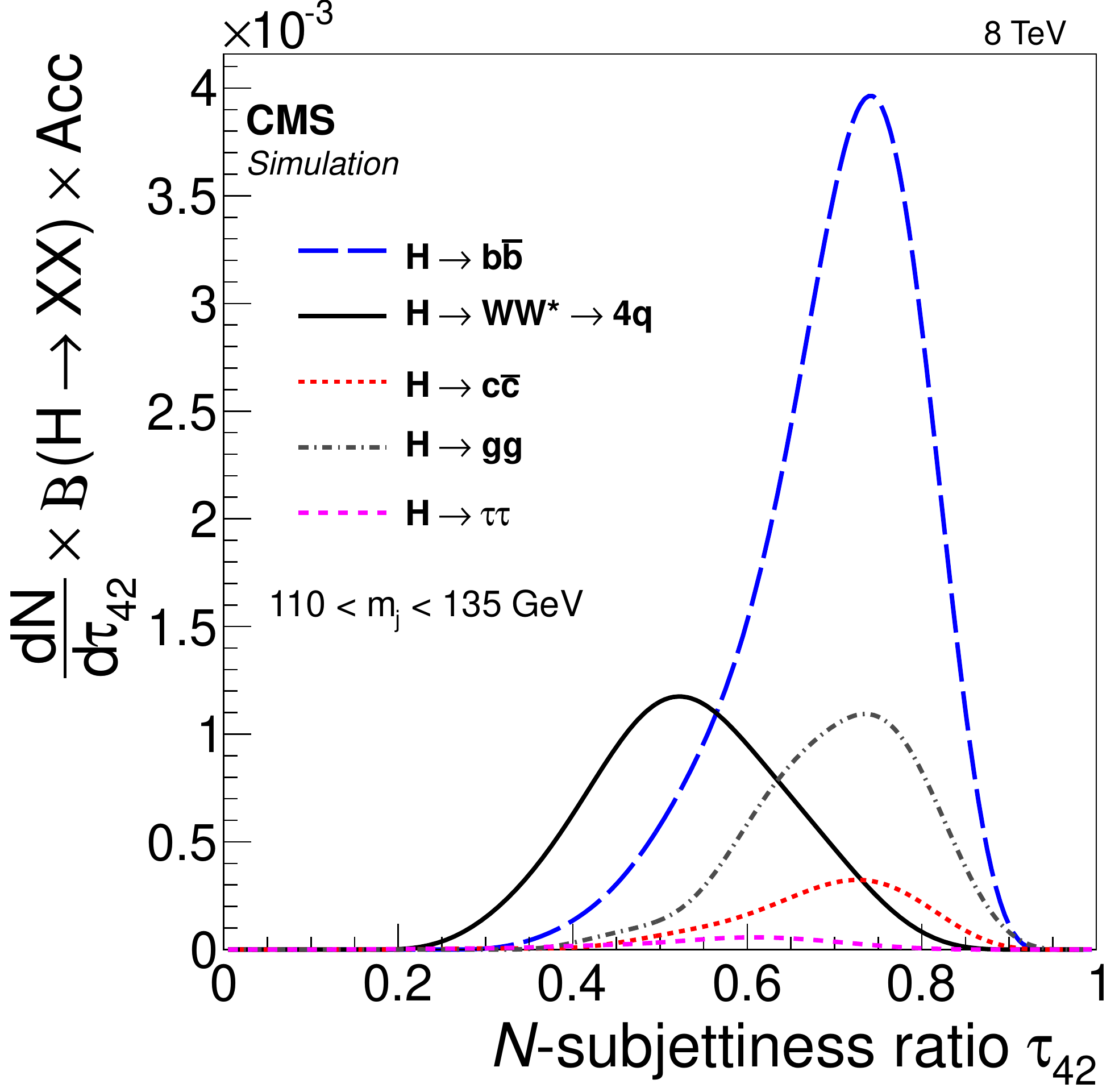}} \\
\end{tabular}
\caption{Comparison of $\tau_{42}$ distributions
for signal events failing the $\Hbb$ requirement.
These events are from the $\Hww$, $\Hbb$, $\PH\to\Pg\Pg$,
$\PH\to\PQc\PAQc$, and $\PH\to\tau\tau$ channels.
The H jets are from a 1.5\TeV resonance decaying to VH.
 All curves are normalized to the product of the corresponding branching
fraction and acceptance.}
\label{fig:HiggsCTalk}
\end{figure}

Cross-talk between the H decay channels is possible;
for example, two-pronged H decays
(e.g. $\Hbb$, $\PH\to\PQc\PAQc$) can be reconstructed as four-pronged $\Hww$, as shown in Fig.~\ref{fig:HiggsCTalk}.
Because of its large branching fraction, $\Hbb$ contributes
a non-negligible number of events to the $\Hww$ tagged sample.
In order to combine events from $\Hbb$ and $\Hww$ channels
into a single joint
likelihood, these categories must be mutually exclusive.
Since the $\Hbb$ tagger has significantly lower background than $\Hww$,
it takes precedence in selecting events. We first identify the
events that pass the $\Hbb$ tagger, and only if they fail we
test them for the presence of the $\Hww$ tag.
Thus we arrive at the final division of events into five mutually
exclusive categories.
These event categories and their nomenclature are
summarized in Table~\ref{table:categories}.

\begin{table}[htb]
\centering
  \topcaption{
        Summary of event categories and their nomenclature used in the paper.
The jet mass cut is $70 < m_\mathrm{j} < 100\GeVcc$
for the V tag and $110 < m_\mathrm{j} < 135\GeVcc$ for the H tag.
    \label{table:categories}}
\begin{tabular}{ ccc}
\hline
\setlength{\tabcolsep}{24pt}
Categories                & V tag                 & H tag               \\
\hline
\rule{0pt}{2.4ex} \HbbHP\ & $ \tau_{21} \leq 0.5$    & b tag              \\
\rule{0pt}{2.4ex} \HbbLP\ & $ 0.5 <\tau_{21} < 0.75$ & b tag              \\
\rule{0pt}{2.4ex} \HWWHP\  & $\tau_{21}\leq0.5$ & $\tau_{42} \leq 0.55$ \\
\rule{0pt}{2.4ex} \HWWLPV\ & $0.5<\tau_{21}< 0.75$ & $\tau_{42} \leq 0.55$ \\
\rule{0pt}{2.4ex} \HWWLPH\ & $\tau_{21}\leq 0.5$ & $0.55 < \tau_{42} < 0.65$ \\
\hline
\end{tabular}
\end{table}

The LP V tag and LP H tag category is not included in this analysis,
since it is dominated by background and therefore
 its contribution to the expected significance of the signal is negligible.
Other H decay modes like $\PH\to\Pg\Pg$, $\PH\to\tau\tau$,
$\PH\to\Z\Z^*$,
and $\PH\to\PQc\PAQc$ together
contribute 2--7\%
of the total $\Hbb$ tagged events, and 18--24\%
of the total $\Hww$ tagged events, as shown in Fig.~\ref{fig:HiggsCTalk}.
In this analysis, we only consider the $\Hbb$ and $\Hww$ channels.
Other H channels passing the
tagging requirements are conservatively viewed as background and
included as systematic uncertainties,
discussed in Section~\ref{sec:systematics}.

The expected tag probabilities of the \PW, Z, and H selection
criteria for
signal and data events in different event categories are shown
in Figs.~\ref{fig:HbbZqqOverallEff} and~\ref{fig:HwwEffAll},
 as a function of $m_\mathrm{jj}$.
The $\PW/\cPZ$ and $\Hww$ tagging efficiencies
for signal events in the $\Hww$ categories
fall at high \pt,
primarily because the $\tau_{42}$
 distribution is \pt-dependent.

The Monte Carlo modelling of V-tag efficiency is validated
using high-\pt $\PW\to\cPaq \Pq'$
decays selected from a data sample enriched
in semileptonic $\ttbar$ events~\cite{JME-13-006}.
Scale factors of \scalefactorHP~and
\scalefactorLP are
applied to the simulated events in the HP and LP V tag categories,
respectively,
to match the tagging efficiencies
 in the top pair data.
The decay of $\Hww$ produces a hard W jet accompanied by two soft jets
from the off-shell W boson.
As the $\Hww$ tagger is also based
on the $N$-subjettiness variables,
and the measured ratio $\tau_{42}/\tau_{21}$
is well modelled by QCD simulation, it
is reasonable to assume that the mismodelling
of the shower by \PYTHIA is similar to that
in the case of V tagging.
The $\Hww$ tagging efficiency scale factors are extrapolated
using the same technique as
for V tagging for both the HP and LP categories, respectively, with
additional systematic uncertainties, which are discussed in Section~\ref{sec:systematics}.

\section{Resonance search in the dijet mass spectrum}
 \label{sec:background}

The resolution for the $m_\mathrm{jj}$ reconstruction is
in the range 5--10\% for all the five categories.
The dominant background in this analysis is from multijet events with
an additional 1--3\% contribution from $\ttbar$ events.
The background is modelled by a smoothly falling
distribution for each event category, given by the empirical
probability density function
\begin{equation}
P_D(m_\mathrm{jj}) = \frac{P_{0} (1 - m_\mathrm{jj}/\sqrt{s})^{P_{1}}}{(m_\mathrm{jj}/\sqrt{s})^{P_{2}}}.
\label{eqParam}
\end{equation}
\noindent
The background model includes the small $\ttbar$ background,
which falls smoothly in a similar way to the multijet background.

 Each event category has separate normalization $P_0$ and shape parameters $P_1$
and $P_2$.
This
parameterization was deployed successfully in a number of
searches based on dijet mass
spectra~\cite{cmsdijet}. A Fisher F-test~\cite{Ftest} is used to check
that no additional parameters are needed to model the individual
background distributions, compared with the four-parameter function used in~\cite{cmsdijet}.
 We have also tested an alternative function $P_E(m_\mathrm{jj}) = P_{0}/{(m_\mathrm{jj}/\sqrt{s} + P_{1})}^{P_{2}}$,
and found it less favored by the F-test.

The use of the alternative function in
the analysis produces negligible changes
in the final result and therefore, no systematic
uncertainty is associated with
this choice.

We search for a peak on top of the falling background spectrum by
means of a binned maximum likelihood fit to the data.

The binned likelihood is given by
\begin{equation}
\mathcal{L} = \prod_{i}\frac{\lambda_{i}^{n_{i}} \re^{-\lambda_{i}}}{n_{i}!},
\end{equation}
where ${\lambda_{i}} = {\mu}{N_{i}(S)} + {N_{i}(B)}$,
$\mu$ is a scale factor for the signal, $N_i(S)$ is the number
of events expected from the signal,
and $N_i(B)$ is the number expected from
multijet background. The variable
$n_i$ quantifies the number of observed
events in the $i^\mathrm{th}$ $m_\mathrm{jj}$ bin.
The number of background events
$N_i(B)$ is described by the functional form of
Eq.~(\ref{eqParam}).
The signal shape for each narrow-width resonance hypothesis is
obtained by fitting the  $m_\mathrm{jj}$ distribution from simulated events  
with a sum of a Gaussian
and a Crystal Ball probability density function.  The resulting
shape is fixed and, as such, used in the 
combined signal and background fit.  This procedure is 
repeated for each resonance hypothesis, sampling resonance masses
from 1.0 to 2.6\TeV in steps of 50\GeV. 
While maximizing the likelihood, $\mu$ and
the parameters of the background function are left unconstrained.
The shape of the resonance is additionally modified to
account for systematic uncertainties (described below); parameters
controlling each source of systematic uncertainty are also allowed
to vary in
the fit, albeit within constraints.
For presentational purposes, a binning according to $m_\mathrm{jj}$ resolution
is used in this paper.  However, the likelihood is calculated
in bins of 1\GeV in $m_\mathrm{jj}$,
approximating an unbinned analysis,
while keeping it computationally manageable.

\begin{figure}[th!b]
\begin{center}
\includegraphics[width=0.49\textwidth]{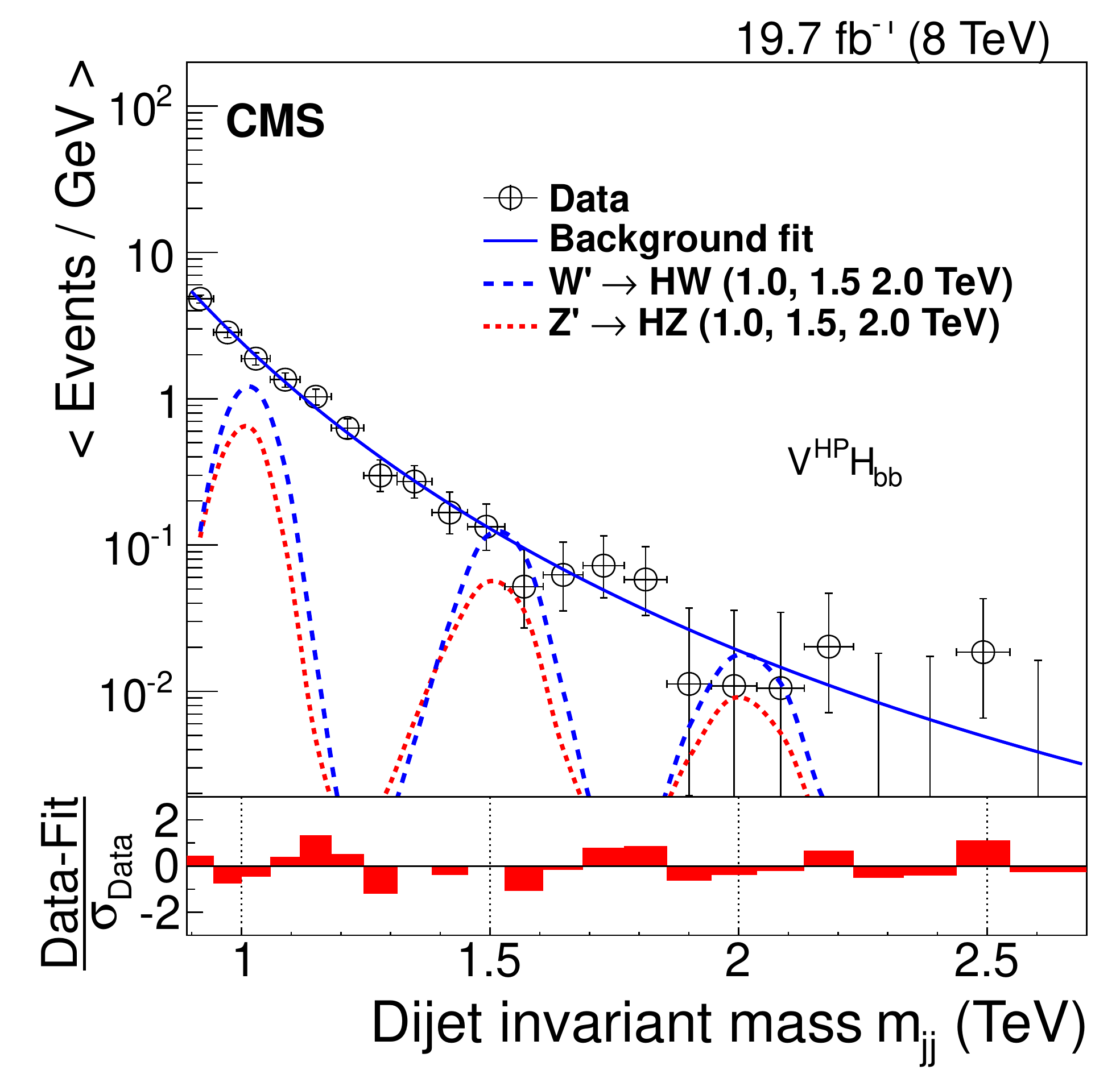}
\includegraphics[width=0.49\textwidth]{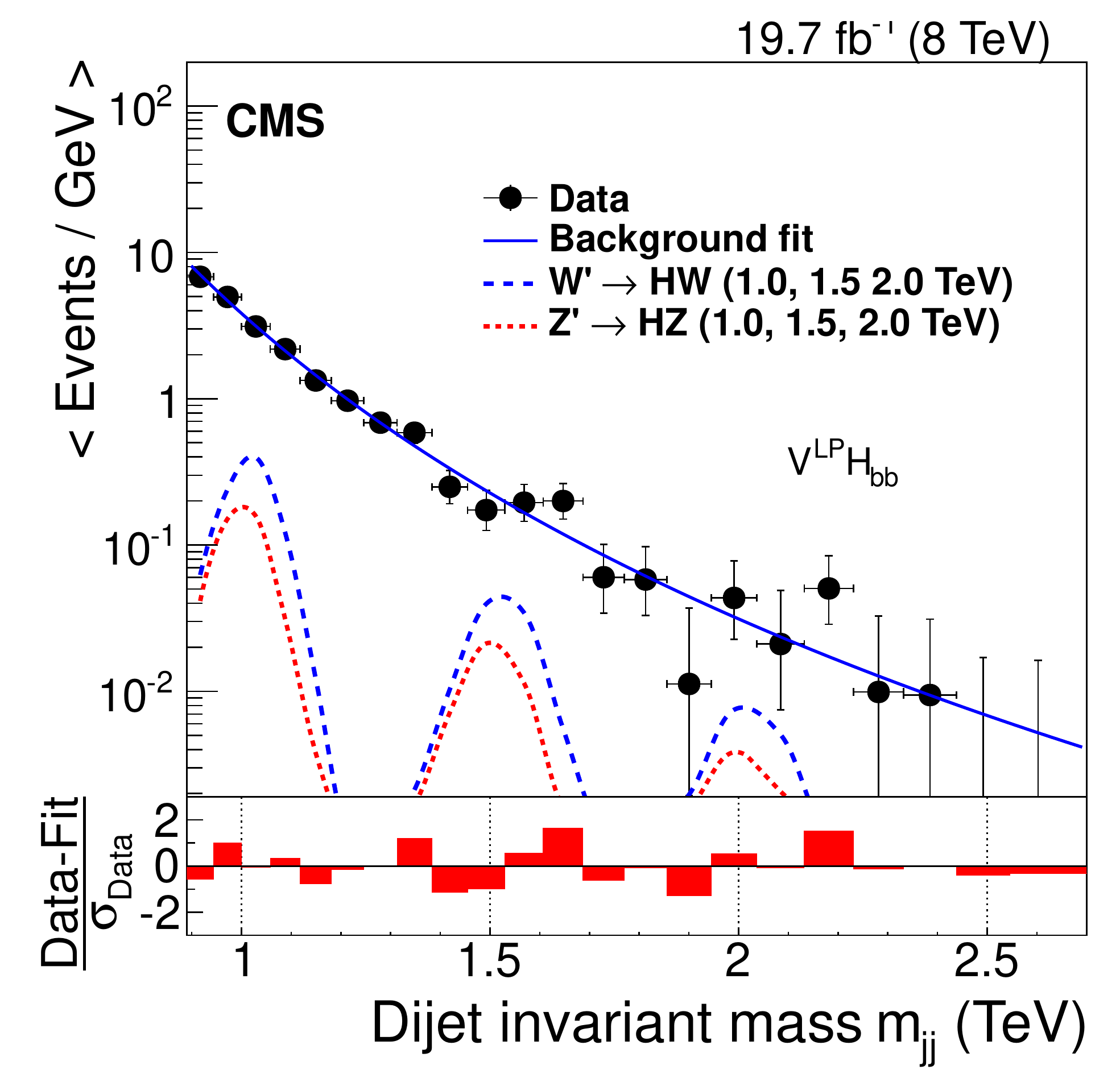}
\end{center}
\caption{Distributions in $m_\mathrm{jj}$ are shown for
   \HbbHP\ category (left), \HbbLP\ category (right).
    The solid curves represent the
   results of fitting Eq.~(\ref{eqParam}) to the data. The
   distributions for \HbbVqq\
   contributions, scaled to their corresponding cross sections, are
   given by the dashed curves.
   The vertical axis displays the number of events per bin, divided
   by the bin width.
   Horizontal bars
   through the data points indicate the bin width.
   The corresponding pull
   distributions
   $\frac{\text{Data}-\text{Fit}}{\sigma_{\text{Data}}}$, where
   $\sigma_{\text{Data}}$ represents the statistical uncertainty in
   the data in a bin in $m_\mathrm{jj}$, are shown below each
   $m_\mathrm{jj}$ plot.}
\label{fig:HbbZqqBG}
\end{figure}

\begin{figure}[th!b]
\begin{center}
\includegraphics[width=0.51\textwidth]{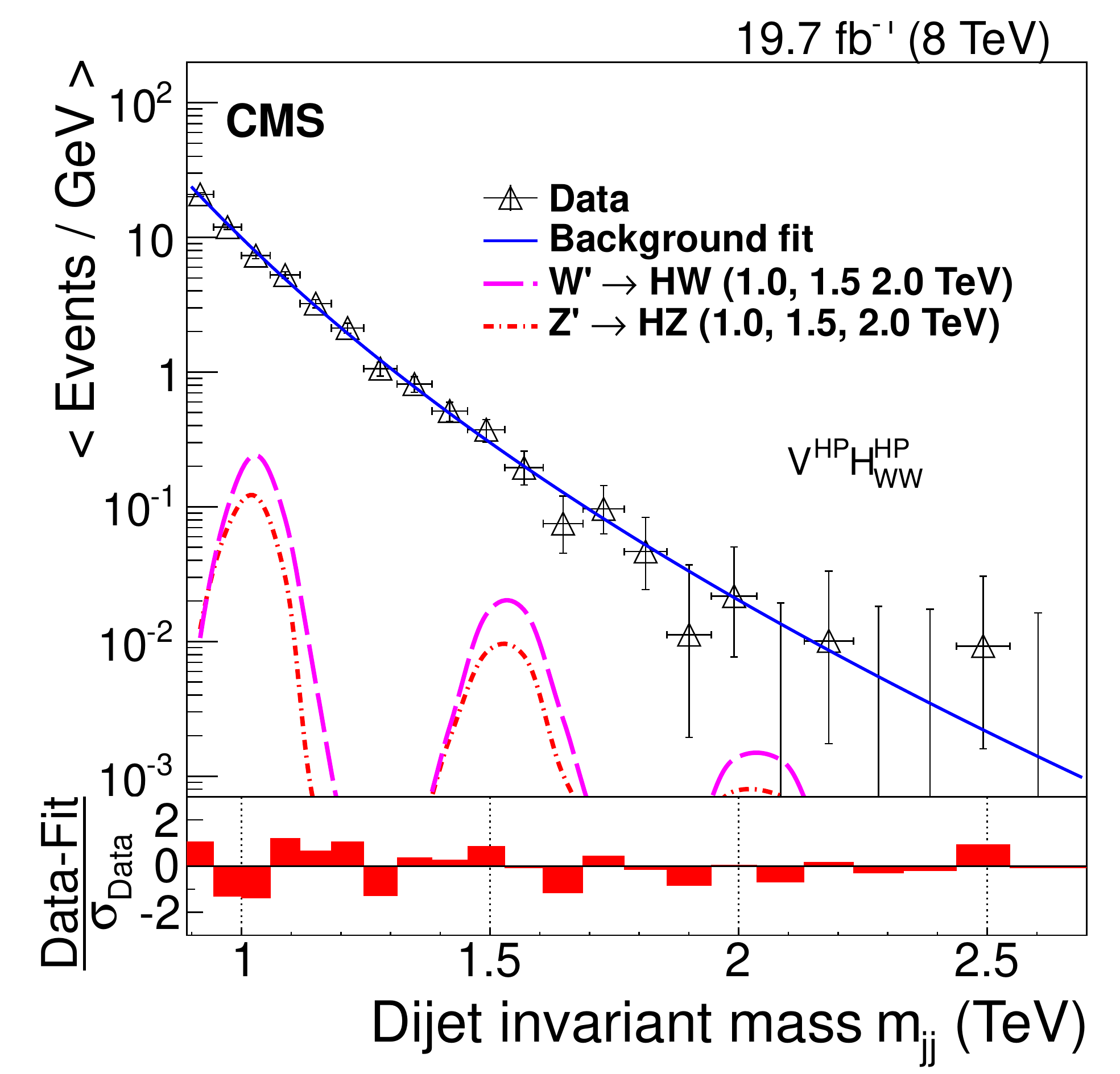}
\includegraphics[width=0.49\textwidth]{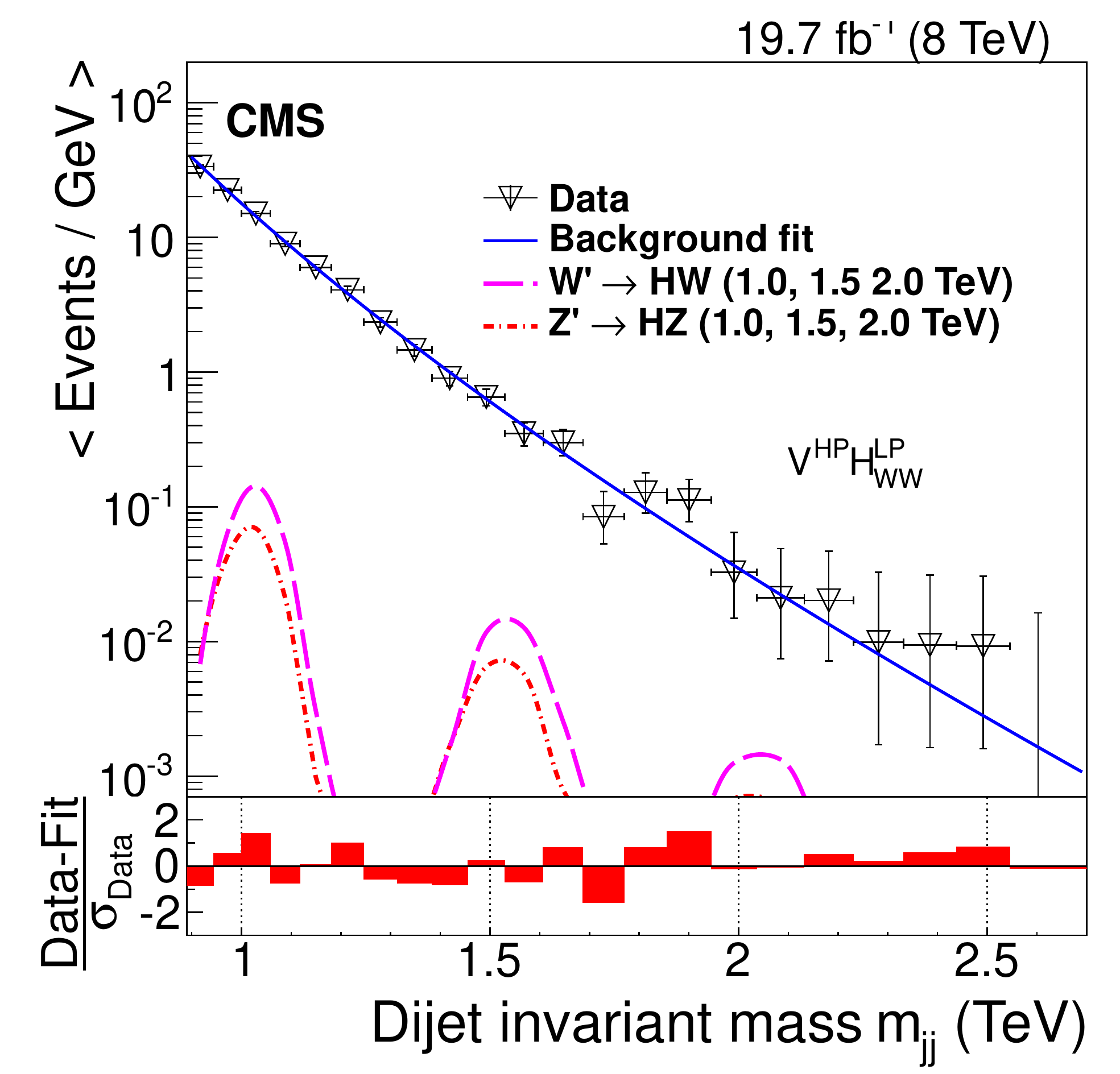}
\includegraphics[width=0.49\textwidth]{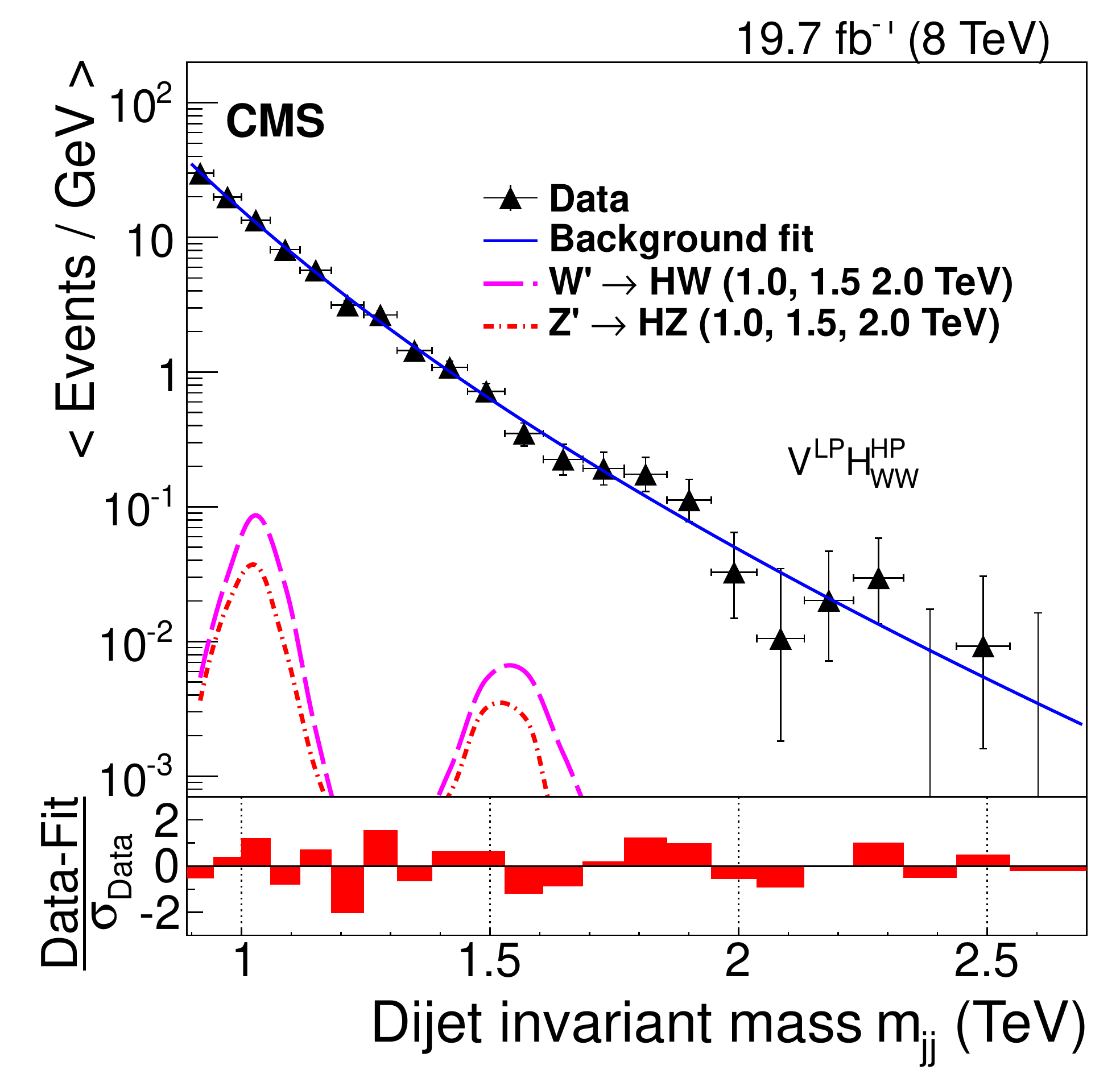}
\end{center}
\caption{
Distributions in $m_\mathrm{jj}$ are shown for
   \HWWHP\ (top), \HWWLPH\ (bottom left), and
   \HWWLPV\ (bottom right).
    The solid curves represent the
   results of fitting Eq.~(\ref{eqParam}) to the data. The
   distributions for \HwwVqq
   contributions, scaled to their corresponding cross sections, are
   given by the dashed and dash-dotted curves.
   The vertical axis displays the number of events per bin, divided
   by the bin width.
Horizontal bars
through the data points indicate the bin width.
The corresponding pull
   distributions
   $\frac{\text{Data}-\text{Fit}}{\sigma_{\text{Data}}}$, where
   $\sigma_{\text{Data}}$ represents the statistical uncertainty in
   the data in a bin in $m_\mathrm{jj}$, are shown below each
   $m_\mathrm{jj}$ plot. }
\label{fig:HwwZqqBG}
\end{figure}

Figs.~\ref{fig:HbbZqqBG} and~\ref{fig:HwwZqqBG}
show the $m_\mathrm{jj}$ distributions in data.
The solid
curves represent the results of the maximum likelihood fit to the data,
fixing the number of expected signal events to zero, while the bottom
panels show the corresponding pull distributions, quantifying the
agreement between the background-only hypothesis and the data.
The expected distributions of \HbbVqq~and \HwwVqq~signals at
1.0, 1.5 and 2.0 \TeV in each category,
scaled to their corresponding cross sections
are given by the dashed and dash-dotted
curves.
The resonance masses in \HbbAll\ channels are slightly lower than
those of the \HWWAll\ channels because of missing
neutrinos in b-hadron decays
and partial misreconstruction of two-pronged ${\rm\Hbb}$ decays.

\clearpage
\newpage

\section{Systematic uncertainties}
\label{sec:systematics}

The largest contributions to
the systematic uncertainty are
associated with the modelling of the signal,
namely: the efficiencies of $\PW/\cPZ$, H, and b tagging;
the choice of PDF; the jet energy
scale (JES); the jet energy resolution (JER); the
pileup corrections; the cross-talk between
different signal contributions; and the integrated luminosity.

The uncertainty in the efficiency for $\PW/\cPZ$ tagging
is estimated using a control sample enriched with $\ttbar$
events
described in Ref.~\cite{JME-13-006}.
Uncertainties of\scalefactorHPu~and\scalefactorLPu~in
 the respective scale factors for HP and LP V tag
include contributions from control-sample statistical uncertainties,
and the uncertainties in the JES and JER for
pruned jets~\cite{Khachatryan:2014hpa}.
The uncertainty due to the extrapolation of the simulated
$\PW/\cPZ$-tagging efficiency to higher jet \pt is
estimated by studying the $\PW/\cPZ$-tagging efficiency as a function of
\pt for two different showering and hadronization models using
\PYTHIA~6 and \HERWIG{++}, respectively.
The results show that the differences are within $4\%$ ($12\%$)
for the HP (LP) V tagging~\cite{JME-13-006}.

We extrapolate the $\Hww$ tagging efficiency
scale factor in the same way as the W/Z-tagging efficiency, with
an additional systematic uncertainty based on the difference between
\PYTHIA~6 and \HERWIG{++} in modelling $\Hww$ decay.
This is evaluated to be ${\approx}$7\%
for the HP and LP H tag.
The uncertainty from the pruned jet mass requirement in the $\Hww$
search is already included in the extrapolated
scale factor uncertainty of the V-tag.

The uncertainty in the efficiency of $\Hbb$ tagging can be separated into two categories: the
efficiency related to the b tagging and the efficiency related to the pruned H mass tag.
The first is
obtained by varying the b-tagging scale factors within the associated uncertainties~\cite{BTV-13-001}
and amounts
to 15\%. The second is assumed
to be similar to the mass selection efficiency of
W jets estimated in Ref.~\cite{JME-13-006}, additionally accounting
for the difference in fragmentation of light quarks and
b quarks, which amounts to 2.6\% per jet.

Because of the rejection of charged
particles not originating from the primary vertex, and the application
of pruning, the dependence of the W/Z- and H-tagging
efficiencies on
pileup is weak and the uncertainty in the modelling of the pileup
distribution is $\leq$1.5\% per jet.

In this analysis, we only consider $\Hbb$ and $\Hww$ decays.
Other H decay channels that pass H taggers are viewed
 as nuisance signals,
and a corresponding cross-talk systematic uncertainty is assigned.
We evaluate this uncertainty as a ratio of expected nuisance signal
events
with respect to the
 total expected signal events, taking into account
the branching fractions, acceptances and tagging efficiencies.
The contamination from cross-talk is estimated to be 2--7\%
in the \HbbAll\ categories, and 18--24\% in the \HWWAll\ categories, and
we take the maximum as the uncertainty.
The analysis is potentially 7\% (24\%) more sensitive than quoted,
 but since it is not clear how well the efficiency for
 the nuisance signals is understood,
 they are neglected, yielding a conservative limit on new physics.
When the \HbbAll\ and \HWWAll\ categories are combined together,
the 24\% uncertainty
becomes a small effect, based on a quantitative measure of
sensitivity
 suggested in Ref.~\cite{punzi}:
\begin{equation}
P = \frac{\mathcal{B}(\PH \to \mathrm{XX}) \, \epsilon_{S}}{1+\sqrt{N_\mathrm{B}}}
\end{equation}
where ${\mathcal{B}(\PH \to \mathrm{XX})}$ is the branching fraction for
the H decay channel, $\epsilon_{S}$ is the signal tagging efficiency,
and $N_B$ is the corresponding background
yield.
The values of $P$ for each channel are shown in Table~\ref{table:SB}.

\begin{table}[htb]
\centering
  \topcaption{
     Summary of the values $P$ for a $\cPZpr$ signal at 1.5\TeV resonance mass
    and the corresponding background yield
    in all five categories.
    \label{table:SB}}
\begin{tabular}{ ccccccc}
\hline
\rule{0pt}{2.4ex} Signal/Categories  & \HbbHP\ & \HbbLP\ & \HWWHP\ & \HWWLPH\ & \HWWLPV\ \\
\hline
\rule{0pt}{2.4ex} \HbbZqq\  & 2.3 $\times 10^{-2}$       &   4.8 $\times 10^{-3}$     &   1.0 $\times 10^{-3}$     & 1.6 $\times 10^{-3}$  & 3.9 $\times 10^{-4}$   \\
\HwwZqq\  & 5.6 $\times 10^{-4}$ & ${\approx}0$  &  2.6 $\times 10^{-3}$  & 9.8 $\times 10^{-4}$ &  4.5 $\times 10^{-4}$     \\
\hline
\end{tabular}
\end{table}

The JES has an uncertainty of
1--2\%~\cite{JME-JINST,Collaboration:2013dp}, and its \pt and $\eta$
dependence is propagated to the reconstructed value of
$m_\mathrm{jj}$, yielding an uncertainty of 1\%, independent
of the resonance mass. The impact of this uncertainty on the
calculated limits is estimated by changing the dijet mass in the
analysis within its uncertainty. The JER is known to a precision of
10\%, and its non-Gaussian features observed in data are well
described by the CMS simulation~\cite{JME-JINST}. The effect of the
JER uncertainty on the limits is estimated by changing the
reconstructed resonance width within its uncertainty. The integrated
luminosity has an uncertainty of 2.6\%~\cite{LUM-13-001}, which is
also taken into account in the analysis.

 The uncertainty related to
the PDF used to model the signal acceptance is estimated from
the CT10~\cite{ct10}, MSTW08~\cite{mstw08}, and
NNPDF21~\cite{NNPDF} PDF sets.
The envelope of the upward and downward variations of the
estimated acceptance for the three sets is assigned as uncertainty~\cite{Alekhin:2011sk}
and
found to be 5--15\% in the resonance mass range of interest. A
summary of all systematic uncertainties is given in
Table~\ref{table:systematicsAll} and~\ref{table:systematicsAllCate}.
Among these uncertainties, the JES
and JER are applied as shape uncertainties, while others
are applied as uncertainty in the event yield.

\begin{table}[htbp]
\topcaption{
Systematic uncertainties common to all categories.
}
\centering
\begin{tabular}{ccc}
\hline
Source &  HP uncertainties (\%)   & LP uncertainties (\%) \\  \hline
JES       &   1     &   1 \\
JER   & 10      & 10   \\
Pileup              & $\leq$3.0     &  $\leq$3.0    \\
PDF                     & 5--15   & 5--15  \\
Integrated luminosity   & 2.6     &  2.6  \\
W tagging               & 7.5     &  54 \\
W tag \pt dependence     & 4       & 12 \\ \hline
\end{tabular}
\label{table:systematicsAll}
\end{table}

\begin{table}[htbp]
\topcaption{
Systematic uncertainties(\%) for $ \mathrm{X} \to \cPV\PH$ signals, in which
$\Hbb$ and  $\Hww$.
Numbers in parentheses represent the
uncertainty for the corresponding LP category. If LP has the same
uncertainty as HP, only the HP uncertainty is presented here.}
\centering
\begin{tabular}{cccc}
\hline
         & \multicolumn{3}{c}{Final state} \\
 Source      \rule{0pt}{2.2ex}   &    \multicolumn{2}{c}{$\Hbb$} &\multicolumn{1}{c}{$\Hww$}    \\
          &  \HbbAll\   & \multicolumn{1}{c}{\HWWAll\ }   & \HWWAll\  \\  \hline

\rule{0pt}{2.4ex} $\Hbb$ mass scale  & 2.6 & --- & --- \\
H(4q) tagging  & --- & 7.5 (54)            & 7.5 (54) \\
H(4q)-tag $\tau_{42}$ extrapolation & --- & 7 & 7 \\
Cross-talk               & 7 & 24       & 24 \\
b tagging & $\leq 15$ & $\leq 15$  & --- \\ \hline
 \end{tabular}

\label{table:systematicsAllCate}
\end{table}

\section{Results}
\label{sec:results}

The asymptotic approximation~\cite{AsymptCLs} of the LHC
$\mathrm{CL_s}$ criterion~\cite{CLs1,CLs3} is used to set upper limits on
the cross section for resonance production. The dominant sources of
systematic uncertainties are treated as nuisance parameters associated
with log-normal priors in those variables.
For a given value of the
signal cross section, the nuisance parameters are fixed to the values
that maximize the likelihood, a method referred to as
profiling. The dependence of the likelihood on parameters used to
describe the background in Eq.~(\ref{eqParam}) is treated in the same
manner, and no additional systematic uncertainty is assigned
to the parameterization of the background.

\begin{figure*}[ht!pb]
\begin{center}
\includegraphics[width=0.49\textwidth]{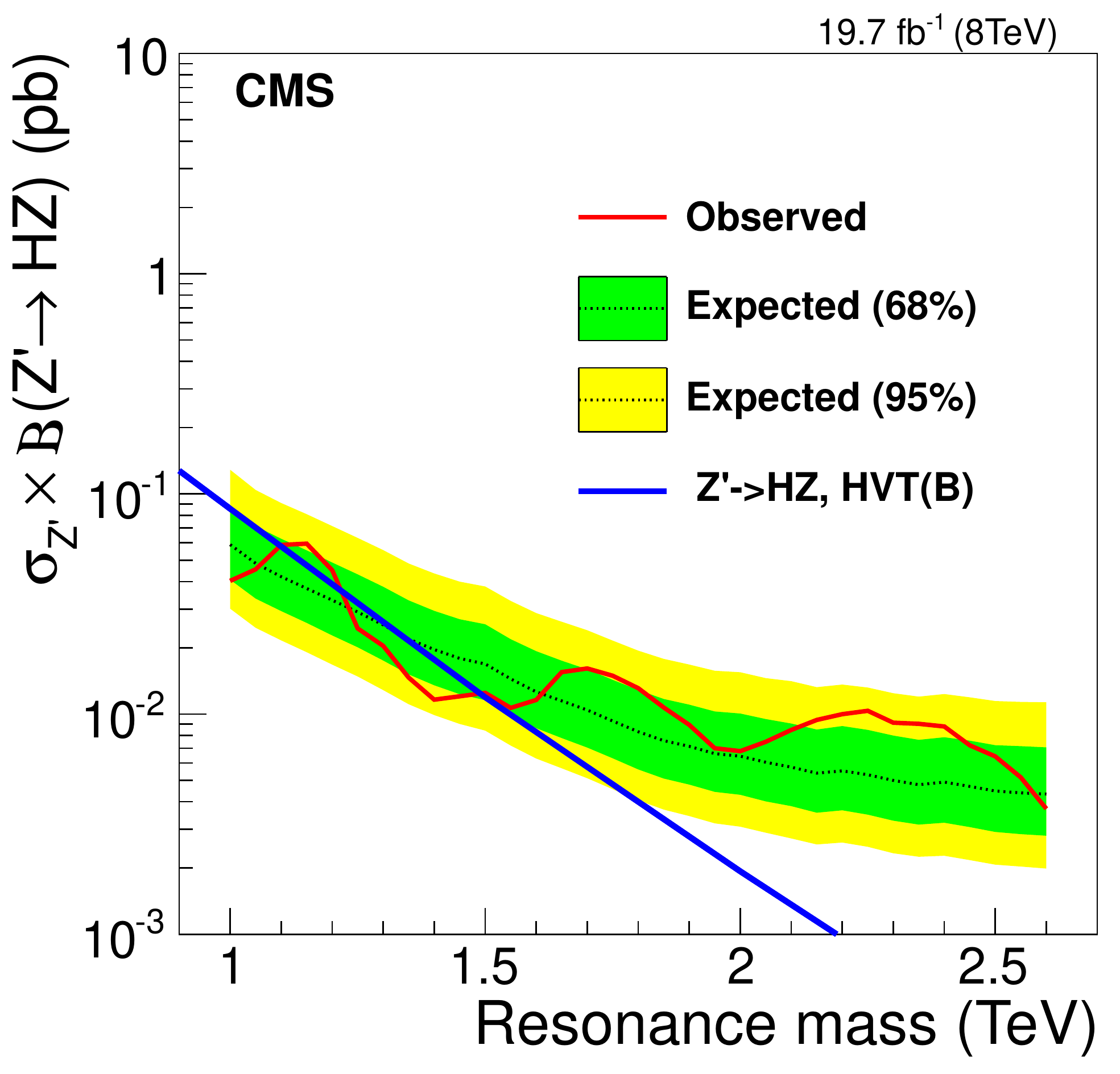}
\includegraphics[width=0.49\textwidth]{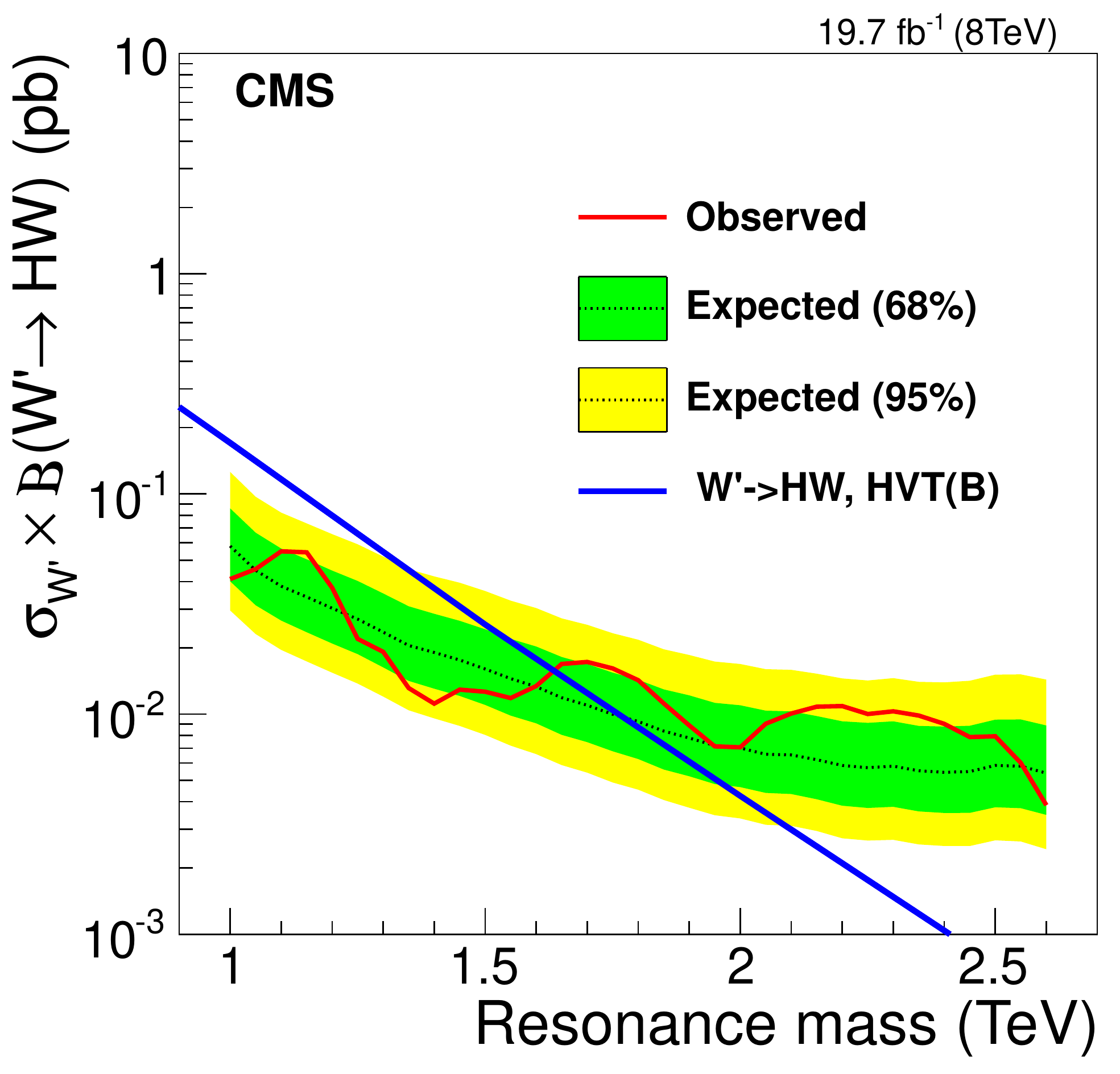}
\end{center}
\caption{Expected and observed upper limits on the production
cross sections
for $\cPZpr\to \PH\Z$ (left) and $\PWpr \to \PH\PW$ (right),
including all five decay categories.
 Branching fractions of H and V decays
have been taken into account.
 The theoretical predictions
of the HVT model scenario B are also shown.
}
\label{fig:HVCombined}
\end{figure*}

\begin{figure*}[ht!pb]
\begin{center}
\includegraphics[width=0.60\textwidth]{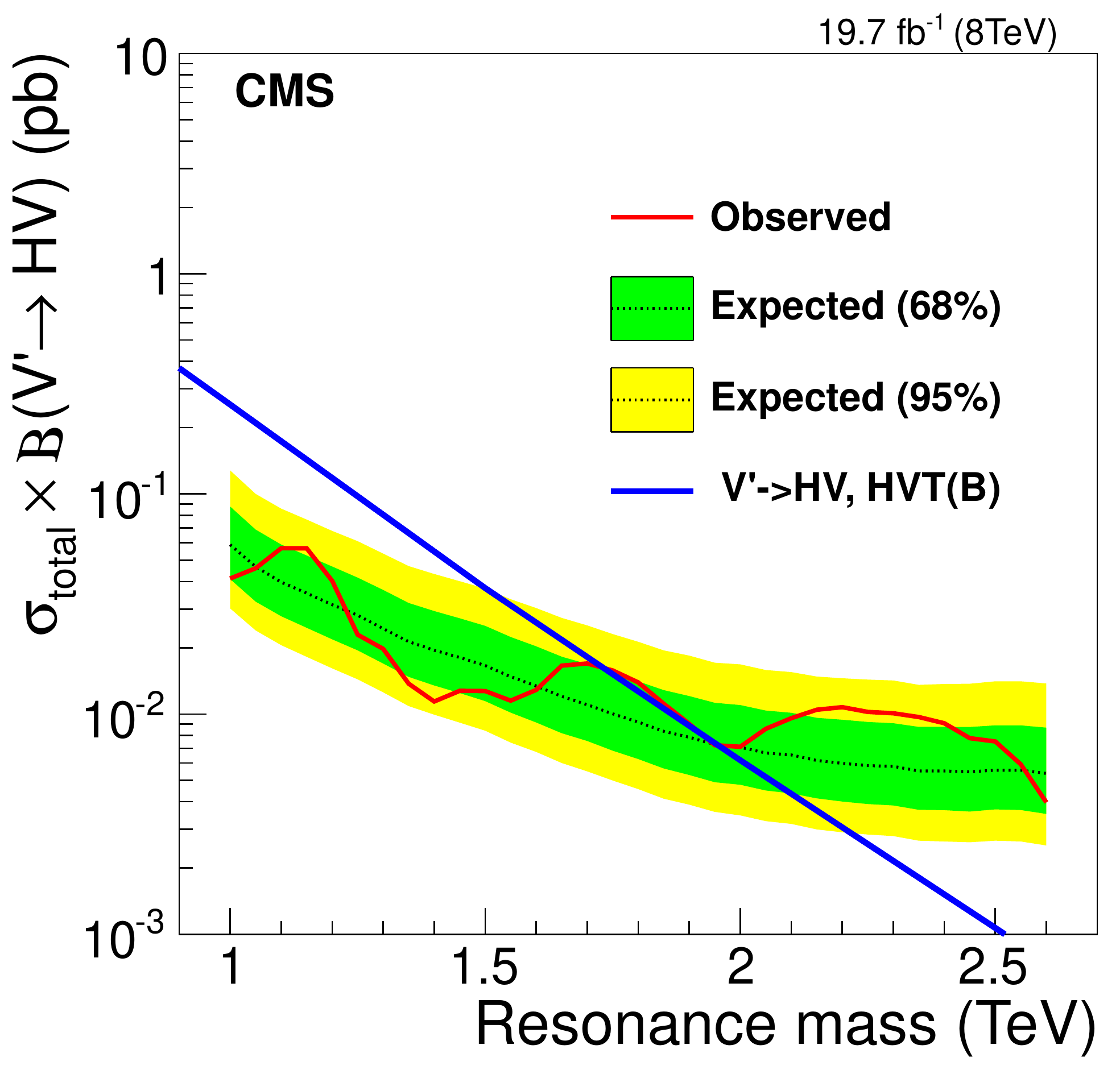}
\end{center}
\caption{Expected and observed upper limits on the production cross section for ${\rm V'\to VH}$,
obtained by combining $\PWpr$ and $\cPZpr$ channels together.
 Branching fractions of H and V decays
have been taken into account.
 The theoretical prediction
of the HVT model scenario B is also shown.
}
\label{fig:HWHZ}
\end{figure*}

 Events from the 5 categories of Table~\ref{table:categories}
are combined into a
 common likelihood, with the uncertainties of the HP and LP H tag (V tag)
efficiencies considered to be anticorrelated between HP
and LP tagging because events failing the
HP $\tau_{42}$ ($\tau_{21}$) selection migrate to the LP category
and the fraction of events failing both HP and LP requirements is
small compared to the HP and LP events.
 The branching fractions of $\Hww$ and $\Hbb$ decays
are taken as  fixed values in joint likelihood.
The remaining systematic uncertainties in the signal
are fully correlated across all channels.  The variables
describing the background uncertainties are treated as uncorrelated.
Fig.~\ref{fig:HVCombined} shows
  the observed and background-only expected upper limits
 on the production cross sections for
$\cPZpr$ and $\PWpr$, including both $\Hbb$ and $\Hww$ decays,
computed at 95\% confidence level (CL), with the predicted
cross sections for the benchmark models overlaid for
comparison.
In the HVT model scenario B,
$\PWpr$ and $\cPZpr$ are degenerate in resonance mass,
thus we compute the limit on their combined cross section under this hypothesis, shown in Fig.~\ref{fig:HWHZ}.
Table~\ref{table:results} shows the
exclusion ranges on resonance masses.

\begin{table}[htb]
\centering
  \topcaption{Summary of observed lower limits on resonance masses at 95\% CL
    and their expected values, assuming a null
    hypothesis. The analysis is sensitive to resonances heavier than 1\TeV.\label{table:results}}
\begin{tabular}{ ccc}
\hline
Process      & Observed & Expected \\
& lower mass limit (\TeVns{}) & lower mass limit (\TeVns{}) \\
\hline
$\PWpr \to \PH\PW$ & [1.0, 1.6] & 1.7  \\
$\cPZpr \to \PH\Z$ & [1.0, 1.1], [1.3, 1.5] & 1.3 \\
 $\mathrm{V}' \to \mathrm{V}\PH$     & [1.0, 1.7] & 1.9 \\
\hline
\end{tabular}
\end{table}

\section{Summary}
\label{sec:conclusions}
A search for a massive resonance decaying
into a standard model-like Higgs boson and a W or Z boson is presented.
A data sample corresponding to an integrated luminosity of \intlumi
collected in proton-proton
collisions at $\sqrt{s}=8$\TeV with the CMS detector
has been used to measure the W/Z and Higgs boson-tagged
dijet mass spectra
using the two highest $\pt$ jets within the pseudorapidity range $\abs{\eta} <
2.5$ and with pseudorapidity separation $\abs{\Delta\eta} < 1.3$.  The QCD
background is suppressed using
 jet substructure tagging techniques,
which identify boosted bosons decaying into hadrons.  In particular,
the mass of pruned jets and the $N$-subjettiness ratios
$\tau_{21}$ and $\tau_{42}$, as well as b tagging applied to the subjets
of the Higgs boson jet, are used to discriminate
against the otherwise overwhelming QCD background.  The remaining QCD
background is estimated from a fit to the dijet mass distributions using a smooth function.
We have searched for the signal as a peak on top of the smoothly
falling QCD background.  No significant signal is observed.
In the HVT model B,
a $\cPZpr$ is excluded in resonance mass intervals [1.0, 1.1] and
[1.3, 1.5]\TeV, while a $\PWpr$ is excluded in the interval [1.0, 1.6]\TeV.
A mass degenerate
$\PWpr$ plus $\cPZpr$ particle is excluded in the interval [1.0, 1.7]\TeV.

This is the first search for heavy resonances decaying into
a Higgs boson and a vector boson (W/Z)
 resulting in a hadronic final state,
as well as the first application of jet substructure techniques
to identify
$\Hww$ decays of the Higgs boson at high Lorentz boost.

\newpage

\begin{acknowledgments}
We congratulate our colleagues in the CERN accelerator departments for the excellent performance of the LHC and thank the technical and administrative staffs at CERN and at other CMS institutes for their contributions to the success of the CMS effort. In addition, we gratefully acknowledge the computing centres and personnel of the Worldwide LHC Computing Grid for delivering so effectively the computing infrastructure essential to our analyses. Finally, we acknowledge the enduring support for the construction and operation of the LHC and the CMS detector provided by the following funding agencies: BMWFW and FWF (Austria); FNRS and FWO (Belgium); CNPq, CAPES, FAPERJ, and FAPESP (Brazil); MES (Bulgaria); CERN; CAS, MoST, and NSFC (China); COLCIENCIAS (Colombia); MSES and CSF (Croatia); RPF (Cyprus); MoER, ERC IUT and ERDF (Estonia); Academy of Finland, MEC, and HIP (Finland); CEA and CNRS/IN2P3 (France); BMBF, DFG, and HGF (Germany); GSRT (Greece); OTKA and NIH (Hungary); DAE and DST (India); IPM (Iran); SFI (Ireland); INFN (Italy); MSIP and NRF (Republic of Korea); LAS (Lithuania); MOE and UM (Malaysia); CINVESTAV, CONACYT, SEP, and UASLP-FAI (Mexico); MBIE (New Zealand); PAEC (Pakistan); MSHE and NSC (Poland); FCT (Portugal); JINR (Dubna); MON, RosAtom, RAS and RFBR (Russia); MESTD (Serbia); SEIDI and CPAN (Spain); Swiss Funding Agencies (Switzerland); MST (Taipei); ThEPCenter, IPST, STAR and NSTDA (Thailand); TUBITAK and TAEK (Turkey); NASU and SFFR (Ukraine); STFC (United Kingdom); DOE and NSF (USA).

Individuals have received support from the Marie-Curie programme and the European Research Council and EPLANET (European Union); the Leventis Foundation; the A. P. Sloan Foundation; the Alexander von Humboldt Foundation; the Belgian Federal Science Policy Office; the Fonds pour la Formation \`a la Recherche dans l'Industrie et dans l'Agriculture (FRIA-Belgium); the Agentschap voor Innovatie door Wetenschap en Technologie (IWT-Belgium); the Ministry of Education, Youth and Sports (MEYS) of the Czech Republic; the Council of Science and Industrial Research, India; the HOMING PLUS programme of the Foundation for Polish Science, cofinanced from European Union, Regional Development Fund; the Compagnia di San Paolo (Torino); the Consorzio per la Fisica (Trieste); MIUR project 20108T4XTM (Italy); the Thalis and Aristeia programmes cofinanced by EU-ESF and the Greek NSRF; and the National Priorities Research Program by Qatar National Research Fund.
\end{acknowledgments}

\bibliography{auto_generated}

\cleardoublepage \appendix\section{The CMS Collaboration \label{app:collab}}\begin{sloppypar}\hyphenpenalty=5000\widowpenalty=500\clubpenalty=5000\input{EXO-14-009-authorlist.tex}\end{sloppypar}
\end{document}

%% file: EXO-14-009-authorlist.tex
\textbf{Yerevan Physics Institute,  Yerevan,  Armenia}\\*[0pt]
V.~Khachatryan, A.M.~Sirunyan, A.~Tumasyan
\vskip\cmsinstskip
\textbf{Institut f\"{u}r Hochenergiephysik der OeAW,  Wien,  Austria}\\*[0pt]
W.~Adam, E.~Asilar, T.~Bergauer, J.~Brandstetter, M.~Dragicevic, J.~Er\"{o}, M.~Flechl, M.~Friedl, R.~Fr\"{u}hwirth\cmsAuthorMark{1}, V.M.~Ghete, C.~Hartl, N.~H\"{o}rmann, J.~Hrubec, M.~Jeitler\cmsAuthorMark{1}, V.~Kn\"{u}nz, A.~K\"{o}nig, M.~Krammer\cmsAuthorMark{1}, I.~Kr\"{a}tschmer, D.~Liko, I.~Mikulec, D.~Rabady\cmsAuthorMark{2}, B.~Rahbaran, H.~Rohringer, J.~Schieck\cmsAuthorMark{1}, R.~Sch\"{o}fbeck, J.~Strauss, W.~Treberer-Treberspurg, W.~Waltenberger, C.-E.~Wulz\cmsAuthorMark{1}
% ...
\vskip\cmsinstskip
% ...
1:~~Also at Vienna University of Technology, Vienna, Austria\\
2:~~Also at CERN, European Organization for Nuclear Research, Geneva, Switzerland\\
% ...
